\documentclass[default]{sn-jnl}

\usepackage{amsmath}
\usepackage{url}
\usepackage{float}
\usepackage{subcaption}
\usepackage{algorithm}
\usepackage{algpseudocode}
\usepackage{longtable}
\begin{document}

%%Title
\title{Stress state evolution of a cemented granular material subjected to bond dissolution by Discrete Element Modeling}

%%Authors
\author*[1,2]{\fnm{Alexandre} \sur{Sac-Morane}}
\email{alexandre.sacmorane@duke.edu}
%https://orcid.org/0009-0008-5454-8107

\author[1]{\fnm{Hadrien} \sur{Rattez}}
\email{hadrien.rattez@uclouvain.be}
%https://orcid.org/0000-0002-7245-6563

\author[2]{\fnm{Manolis} \sur{Veveakis}}
\email{manolis.veveakis@duke.edu}
%https://orcid.org/0000-0002-4911-6026

\affil[1]{\orgdiv{Institute of Mechanics, Materials and Civil Engineering, UCLouvain}, \orgaddress{Place du Levant 1}, \city{Louvain-la-Neuve}, \postcode{1348}, \country{Belgium}}

\affil[2]{\orgdiv{Multiphysics Geomechanics Lab, Duke University}, \orgaddress{Hudson Hall Annex, Room No. 053A}, \city{Durham}, \postcode{27708}, \state{NC}, \country{USA}}

%%Abstract
\abstract{Rock weathering is a common phenomenon to consider in various engineering applications that involve underground operations, such as underground storage or geothermal energy extraction. This weathering is induced by the circulation of a reacting fluid in the pores of the rock and can induce changes in materials properties and stress state. This work presents discrete elements simulations of the dissolution of sedimentary rocks to study their stress state evolution. The rock is modeled as a cohesive granular material subjected to debonding. Oedometric conditions are considered during the weathering and the evolution of the coefficient of lateral earth pressure $k_0$, a proxy of the stress state, is tracked. In particular, the influence of the degree of cementation, the confining pressure, the initial value of $k_0$ and the history of loading are investigated.
It has been observed that cemented granular materials tend to reach an attractor configuration for the stress state with the increase of the degree of dissolution. $k_0$ aims to reach the $k_0^{attr}$ value, between 0.3 and 0.4, when all the bounds are dissolved, independently of the initial state. Two main mechanisms have been observed to explain this evolution of the interparticles force configurations and thus of the stress state: the collapse of the unstable chain forces (stable only due to the cementation) and the softening of the grains.}

%%Keywords

\maketitle

%%=======================================================%%

%%Highlights

\textbf{Highlights}
\begin{itemize}
   \item A cemented granular material subjected to debonding by dissolution is investigated in oedometric conditions by the discrete element method.
    \item The influence of several parameters on the evolution of the ratio of horizontal to vertical stresses, i.e. the $k_0$ coefficient, is illustrated.
    \item The existence of an attractor configuration for the stress state is observed when the degree of dissolution increases. 
    \item Two main mechanisms can explain our observations: the collapse of the unstable chain force (stable only due to the cementation) and the softening of the grains.
\end{itemize}

\section{Introduction}

To diminish the impact of human activity on climate evolution, one approach involves the utilization of renewable energy sources, such as wind and solar power, to generate electricity.  However, these energy sources are characterized by variability, necessitating the development of strategies for energy storage and utilization. A promising approach involves the transformation of excess energy into $H_2$. The challenge, however, lies in identifying suitable methods to store the produced $H_2$, which must be subsequently extracted and consumed during periods of energy deficiency. It has been posited that underground storage and salt caverns hold considerable promise for this purpose, offering a greater storage capacity than conventional aboveground infrastructure such as pipelines or tanks \cite{Heinemann2021,Lesueur2023}. 
Likewise, geothermal energy appears as an efficient method to produce energy, and it is defined as the injection of a fluid underground to extract the thermal energy of the Earth \cite{McCartney2016,Barbier2002}.
Furthermore, deep aquifers have emerged as a promising medium for the sequestration of greenhouse gases, including $CO_2$ \cite{Kanin2024}. This engineering application, added to the emergence of renewable energy production, participates in the reduction of the impact of human activities on climate change. If $H_2$ storage and geothermal energy production involve a fluid injection and extraction, $CO_2$ sequestration implies only injection, necessitating underground retention. To assess the viability of the $H_2$ storage and $CO_2$ sequestration techniques, various pilot projects have been initiated \cite{Gunnarsson2018,CarbFix2,Hypster,Ilgen2019}. 

The injection of the fluid into a geomaterial reservoir induces a modification of the fluid chemical composition, and thus its reactivity with the surrounding solid phases, activating or enhancing chemical processes \cite{Liteanu2012,Lesueur2020}. The consequences are numerous: settlement at the surface \cite{LeGuen2007,Brzesowsky2014,SacMorane:PFDEM,SacMorane:PFDEMb}, modifications of the mechanical \cite{Lesueur2023,Wang2016,Nova2003} and hydraulic \cite{Hueckel2005,Vallin2013,Lesueur2020b} properties, or threat to the stability of the caprock \cite{Rohmer2016, Manceau2016}. It has even been correlated to earthquake nucleation \cite{Blocher2018,Rattez2020,Rattez2021,Lengline2023}.
These phenomena and consequences observed in the context of the previously described engineering applications are similar to the numerous examples in nature where geomechanics and chemistry appear deeply intertwined, such as weathering \cite{Palmer1973,Vaughan1984}. Even if the kinetics of the natural processes is much slower than the kinetics of the engineering-induced consequences, \emph{Terzaghi} pointed out that long-term phenomena are the main causes of the instabilities rather than single events \cite{Terzaghi1950}, requiring investigation.

The impact of weathering, due to natural processes or engineering applications, on the mechanical behavior of geomaterials has been previously investigated through both experiments \cite{Ciantia2014, Ciantia2014b, Basu2009, Heap2021} and numerical simulations \cite{Nova2003, Stefanou2014, Gajo2015, Loret2002, Buscarnera2012, Hu2013, Hu2019, Wu2025} for sands (unbonded material) and rocks (bonded material) in a range of configurations. However, only a limited number of studies have focused on examining the influence of weathering on the stress state of the material, even if it can induce the material to reach the yield limit \cite{Shin2008}. 

A seminal paper on the evolution of the stress state during weathering was published by \textit{Vaughan and Kwan} in 1984 \cite{Vaughan1984}. In this work, the authors developed a theoretical model to study the evolution of the stress state during the weathering of rocks to form residual soils. More recently, experimental investigations of dissolution have been conducted using soft oedometer tests, which allow the measurement of radial stresses \cite{Kolymbas1993}. These tests have been performed with calcarenite and artificially cemented sand \cite{Castellanza2004}, glass beads mixed with salt grains \cite{Santamarina2009} or carbonate sand \cite{Viswanath2020}. The majority of these studies have focused on non-cohesive granular materials, except \cite{Castellanza2004}, in which the authors observed an increase in radial stress during dissolution by an acid solution in both the artificially cemented silica sand and the fully calcareous soft rock. However, the rate of increase was different in each case, reaching an asymptotic value in both cases.
 
The theoretical and experimental work has been completed through the implementation of numerical studies, predominantly using the Discrete Element Method \cite{Burman1980, OSullivan2011}. It should be noted that, to the best of our knowledge, all existing numerical studies focus on non-cohesive granular materials and model dissolution as a homogeneous decrease in diameter. In \cite{Santamarina2009}, the authors were able to reproduce experimental results in which an initial decrease of $k_0$ was observed. This coefficient reaches the Rankine active pressure coefficient $k_a$ during dissolution. Additionally, the authors noted that $k_0$ may eventually increase to recover its initial value if dissolution continues. In their investigation of the effect of dissolution kinetics \cite{Viswanath2020}, the authors have observed a similar behavior for rapid dissolution, a decrease followed by an increase of $k_0=\sigma_{II}/\sigma_I$, with $\sigma_{I}$ and $\sigma_{II}$ are the vertical and lateral stresses. However, they observed a monotonic decrease for slower dissolution rates.

\begin{figure}[h]
    \centering
    \includegraphics[width=0.8\linewidth]{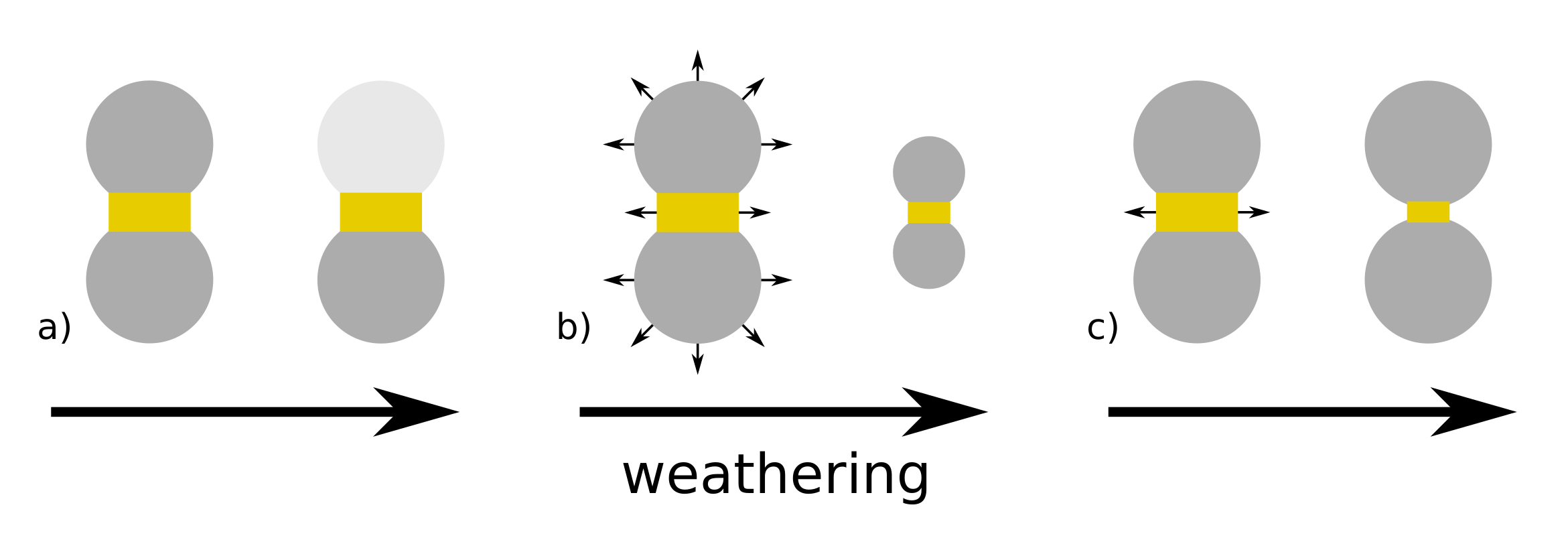}
    \caption{Scheme of the weathering in the context of rock. Three phenomena are distinct: a) the mineral replacement by a weaker phase, b) the dissolution of grains and bonds (calcarenite for example), and c) the dissolution of the bonds, while the grains remain intact (silica sand + carbonate bonds for example).}
    \label{weathering rock}
\end{figure}

The objective of the present study is to investigate the evolution of $k_0$ through numerical simulations solved by Discrete Element Modelization for rock materials. 
In the context of rocks, weathering can be conceptualized as three distinct phenomena depicted in Figure \ref{weathering rock}.
The first chemical alteration is the substitution of a mineral for one with a weaker stiffness or lower density \cite{Detienne2016}.
The second phenomenon is the dissolution of grains and the bonds located between them \cite{Castellanza2004}. 
This degradation is exemplified by calcarenite, wherein the grains and bonds are composed of calcium carbonate.
Indeed, the dissolution kinetics for the bond and the grain appear comparable.  
The third case is the dissolution of the bonds only, while the grains remain intact \cite{Castellanza2004}. An illustrative example is provided by silicic sand, which contains carbonate bonds. The dissolution kinetics of the bond is assumed to be considerably faster than that of the grain. This phenomenon is referred to as debonding and is the subject of the present work.
\emph{Ciantia et al.} have even identified two distinct kinds of bonding within the rock, temporary and persistent bonding, affecting the weathering process \cite{Ciantia2014b}. The temporary bonding origin is identified as the mineral suspension which deposits if the material is dried. In the case of fluid saturation of the sample, these bonds dissolve, with a fast kinetics reaction, as the material is suspended in the pore fluid. Conversely, the persistent bonding origin is the diagenesis phenomenon, and the dissolution kinetics is much smaller. This differentiation between the bonds involves a short and long-term aspect of the mechanical evolution of the geomaterial during weathering. 

Despite the potential of constitutive relations to address chemo-mechanical couplings, these approaches remain constrained to the macroscale.
The phenomena occurring at the microscale, such as granular reorganization and material dissolution, are merely postulated rather than modeled. 
Furthermore, these formulations often consider a homogeneous dissolution within the sample, which is not in agreement with experiments where wormhole formations are observed while injecting acid into a triaxial cell \cite{Xie2011,Algarni2024}. Similarly, chemo-mechanical responses appear localized, and not homogeneous, in the host rock during $CO_2$ sequestration \cite{Ilgen2019}.
The question that remains is \emph{how numerical models which explicitly consider the microstructure predict the mechanical behavior of geomaterials subjected to debonding.}

The Discrete Element Model (DEM) \cite{OSullivan2011}, developed in 1979 by \emph{Cundall and Strack}, has emerged as a pertinent method to predict the mechanical behavior of sands \cite{Burman1980} and rocks \cite{Potyondy2004} at the microscale. In the seminal approach, the granular material is modeled as spheres. The existing bonds in rock are conceptualized as additional stiffnesses between the grains.
The rock material is described as a cohesive granular material following the model introduced in \cite{Bourrier2013}. The mechanical parameters were calibrated by \textit{Sarkis et al.} \cite{Sarkis2022} on biocemented sands \cite{Dadda2017}. Indeed, these biocemented sands can be considered as a suitable proxy for natural sandstones \cite{Konstantinou2021, Wang2023}. While the pressures considered here may not fully reflect the conditions present in an underground reservoir, the observed behavior can be readily extrapolated through an equivalent contact stiffness number ($\propto$ Young modulus/pressure applied) \cite{DaCruz2005, Roux2002}.
This discrete description can be readily extended to capture the chemistry of the weathering. Indeed, the size of the bonds can be reduced to model the silica + carbonate case. These modifications destabilize the granular microstructure and affect the properties at the sample scale.

The present study investigates several parameters, including the degree of cementation, the confining pressure, and the history of load. In particular, the initial value of the coefficient of lateral earth pressure, which serves as a proxy for the state of stress, is monitored and studied. Indeed, this parameter can be affected by tectonic solicitations in the context of underground reservoirs \cite{Hoek2007, Taherynia2016, Demir2018}, reaching a larger value than in experimental setups.

The first Section of this paper is dedicated to the formulation of the model utilized in the Discrete Element Modelization. Subsequently, a second Section describes the framework and the context of this work. Finally, results and discussions are proposed in the next Sections.  

%%=======================================================%%

\section{Theory and formulation}

The Discrete Element Model (DEM) is an approach developed by Cundall \& Strack \cite{Burman1980} for simulating granular materials at the particle level. This model was extended to rock materials by \cite{Potyondy2004}, adding a bond between the particles. The fundamental premise of this methodology is to explicitly consider the individual particles and their interactions within the material itself \cite{OSullivan2011}. Newton's laws (linear and angular momentum) are employed to compute the motion of the grains, which is formulated as follows for a single grain:

\begin{align}
    m\frac{\partial v_i}{\partial t} &= m g_i + f_i \label{Newton Law 1} \\
    I\frac{\partial\omega_i}{\partial t} &= M_i
    \label{Newton Law 2}
\end{align}
where $m$ is the particle mass, $v_i$ is the particle velocity vector, $g_i$ is the gravity acceleration vector, $f_i$ is the sum of contact force vectors applied to the particle, $I$ is the moment of inertia of the particle, $\omega_i$ is the angular velocity vector, $M_i$ is the sum of contact moment vectors applied to the particle (torques due to bending, twisting, and tangential forces).
To solve these equations, various integration methods can be used \cite{Samiei2013}. For example, the software \emph{YADE} used during this work follows a Verlet scheme \cite{YADE}. The time integration depends on a time step $dt$ that must verify the P-wave critical time step condition \cite{Burns2017} defined as:

\begin{equation}
dt<dt_{crit} = \frac{\text{min}(R)}{\sqrt{E_m/\rho}}
\label{Time Step Check}
\end{equation}
where $R$ is the radius of the grains, $E_m$ is the Young modulus of the contact and $\rho$ is the density of the material. Herein, a safety factor is considered, $dt = 0.6\,dt_{crit}$.

Considering two particles with radii $R^1$ and $R^2$, the interaction between particles is computed only if the distance between grains satisfies the following inequality:
\begin{equation}
     \overline{x^1_i-x^2_i}<R^1+R^2
    \label{overlap condition}
\end{equation}
where $x^1_i$ (resp. $x^2_i$) is the center of the particle 1 (resp. 2) and $\overline{u_i}$ is the norm of the vector $u_i$.
Once contact is detected between grains $1$ and $2$, the normal vector of the contact $n^{12}_i$ is computed as $n_i^{12} = (x_i^1-x_i^2)/\overline{x^1_i-x^2_i}$. Then the normal overlap vector $\Delta_{ni}$ and the tangential overlap vector $ \Delta_{si}$ are determined.

\begin{equation}
    \Delta_{ni} = \left(R^1+R^2 - \left(x_j^1-x_j^2\right)n_j^{12}\right)n^{12}_i
\end{equation}

The tangential component $\Delta_{si}$ is computed incrementally, integrating the relative tangential velocity between particles during the contact. 

\begin{equation}
    \Delta_{si} = \Delta_{si} + v_{si}^{12}\times dt
\end{equation}
where $v_{si}^{12}$ is the relative tangential velocity vector defined in Equation \ref{relative tangential velocity} and $dt$ is the time step used in the simulation.

\begin{align}
    v_{si}^{12} &= v^{12}_i-\left(v^{12}_j n^{12}_j\right)n^{12}_i \label{relative tangential velocity}\\
    v^{12}_i&=v_i^1-v_i^2+(R^1-\delta/2)\epsilon_{ijk}n^{12}_j\omega^1_k\nonumber\\
    &+(R^2-\delta/2)\epsilon_{ijk}n^{12}_j\omega^2_k
\end{align}
where $v^{12}_i$ is the relative velocity vector between grains and $\delta = \overline{\Delta_{ni}}$ is the norm of the normal overlap vector. It should be noted that the terms $\left(R-\delta/2\right)$ represent the corrected radii at the contact. Here, the angular velocity vectors $\omega_i$ of the grains are taken into account for the calculation of the tangential overlap vector $\Delta_{si}$. As the contact orientation can evolve, it is crucial to update the tangential overlap vector by rotation and scaling: $\Delta_{si}^{new}=\Delta_{si}^{old}-\Delta_{sj}^{old}n_j^{12}n_i^{12}$ and $\overline{\Delta_{si}^{new}}=\overline{\Delta_{si}^{old}}$.

A relative angular velocity vector $\Delta\omega_i$ is also needed to compute the twisting and bending behaviors.

\begin{equation}
    \Delta \omega_i=\omega^1_i-\omega^2_i 
    \label{Delta omega}
\end{equation}

This relative angular velocity vector $ \Delta\omega_i$ is divided into a twisting component $ \Delta\omega_{ti}$ and into a bending component $\Delta\omega_{bi}$.

\begin{align}
    \Delta\omega_{ti} &=  \Delta\omega_j n^{12}_j n^{12}_i\\
    \Delta\omega_{bi} &= \Delta\omega_i -\Delta\omega_{ti}
\end{align}

These relative angular velocity vectors $\Delta \omega_{ti}$ and $\Delta \omega_{bi}$ are used to compute a twisting and a bending relative angular rotation vectors $\Delta\theta_{ti}$ and $\Delta\theta_{bi}$ incrementally.

\begin{align}
    \Delta\theta_{ti} &= \Delta\theta_{ti} + \Delta\omega_{ti}\times dt\\
    \Delta\omega_{bi} &= \Delta\theta_{bi} + \Delta\omega_{bi}\times dt
\end{align}

As the contact orientation can evolve, it is crucial to update the twisting and bending relative angular rotation vectors by rotation and scaling: $\Delta\theta_{ki}^{new}=\Delta\theta_{ki}^{old}-\Delta\theta_{kj}^{old}n_j^{12}n_i^{12}$ and $\overline{\Delta\theta_{ki}^{new}}=\overline{\Delta\theta_{ki}^{old}}$, with $k=t\text{ or }b$.

\vskip\baselineskip

\begin{figure}[ht]
    \centering
    \includegraphics[width=0.8\linewidth]{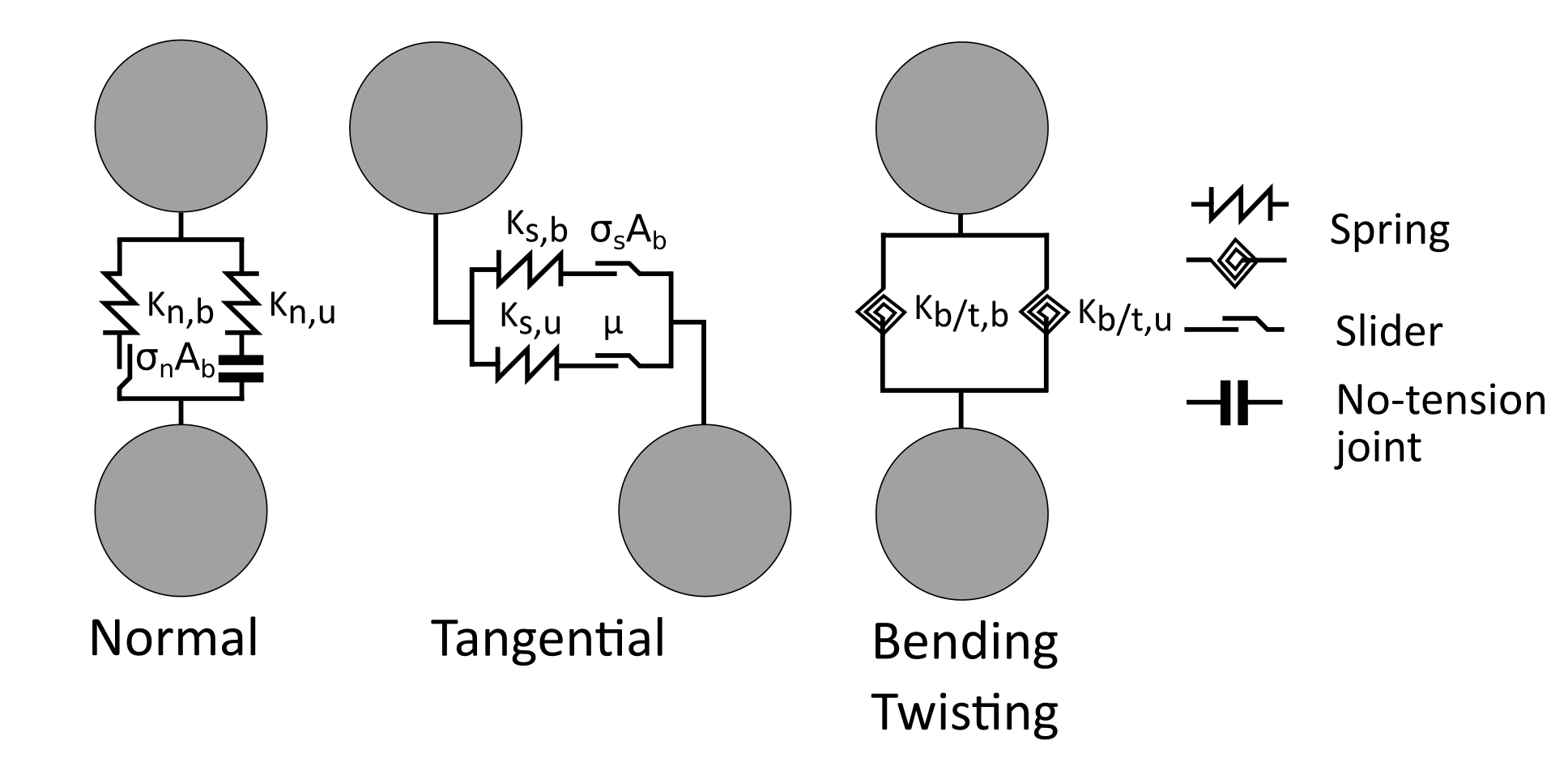}
    \caption{The contact between two particles obeys normal, tangential, bending and twisting elastic-plastic laws. A bond can be present if the contact is cemented. It increases the shear and the tensile strengths.}
    \label{DEM Scheme}
\end{figure}

The contact models between particles considered here are governed by a cohesive law following \cite{Bourrier2013}. This cohesion permits the relative displacement between bonded particles. However, due to the presence of cementation, an additional stiffness is introduced. 
The normal, tangential, bending and twisting models are illustrated in Figure \ref{DEM Scheme} and described in the following. 

\begin{figure}[ht]
    \centering
    \includegraphics[width=0.6\linewidth]{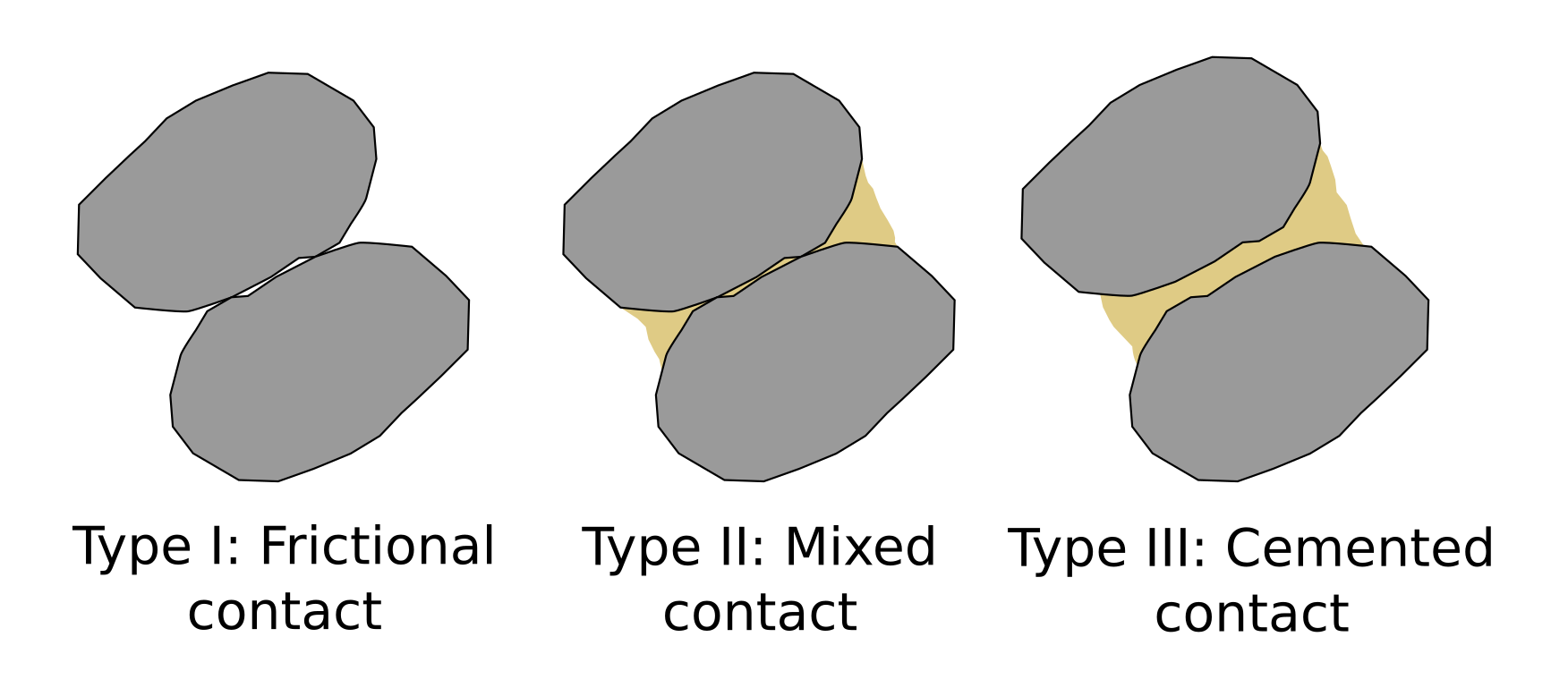}
    \caption{Definition of the contact types: I) Frictional, II) Mixed or cohesive (Frictionnal+Cemented) and III) Cemented.}
    \label{Type contact}
\end{figure}

As illustrated in Figure \ref{Type contact}, a cemented contact can be classified into three distinct types: I) Frictional, II) Mixed/Cohesive (Frictional+Cemented), or III) Cemented \cite{Sarkis2022}. In the case of type I, no cementation is present. In types II and III, the contact can be modeled as two springs in parallel (type II) or in series (type III). One represents the bond $K_{\cdot,b}$ and the other represents the unbonded grain-grain contact $K_{\cdot,u}$.
In the following Sections, the assumption is made that the sample is composed of contacts type II, as depicted in Figure \ref{DEM Scheme} where springs are represented in parallel.

The contact force vector $f_i$ applied to the particle, see Equation \ref{Newton Law 1}, is defined as the sum of the normal force vectors $F_{ni}$ and the tangential force vectors $F_{si}$ over the contacts including the particle. The normal force vector $F_{ni}$ is described in Equation \ref{Normal Force Equation}. An elastic stiffness $K_n=K_{n,b}+K_{n,u}$ is required, formulated Equation \ref{Normal Stiffness Equation}.

\begin{align}
    F_{ni} &= K_n\Delta_{ni} \label{Normal Force Equation}\\
    K_n &= 2E_m\frac{R^1R^2}{R^1+R^2} \label{Normal Stiffness Equation}
\end{align}
where $E_m$ is the contact Young modulus, $R^1$ and $R^2$ are the radii of the particles in stake.

Similarly, the tangential force vector $F_{si}$ is described in Equation \ref{Tangential Force Equation}. An elastic stiffness $K_s= K_{s,b}+K_{s,u}$ is required, formulated Equation \ref{Tangential Stiffness Equation}.
\begin{align}
    F_{si} &= K_s \Delta_{si}\label{Tangential Force Equation}\\
    K_s &= \nu K_n=2\nu E_m\frac{R^1R^2}{R^1+R^2} \label{Tangential Stiffness Equation}
\end{align}
where $\nu$ is the contact Poisson's ratio.

In the same vein, the contact moment vector $M_i$ applied to the particle, see Equation \ref{Newton Law 2}, is defined as the sum of the bending moment vectors $M_{bi}$, the torque due to the tangential force vectors $F_{si}$, and the twisting moment vector $M_{ti}$ over the contacts including the particle.
The bending moment vector $M_{bi}$ is described in Equation \ref{Bending Moment Equation}. An elastic stiffness $K_b=K_{b,b}+K_{b,u}$ is required, formulated Equation \ref{Bending Stiffness Equation}.
\begin{align}
    M_{bi} &= K_b\Delta\theta_{bi}\label{Bending Moment Equation}\\
    K_b &= \alpha_b K_s R^1 R^2=2\alpha_b\nu E_m\frac{\left(R^1R^2\right)^2}{R^1+R^2} \label{Bending Stiffness Equation}
\end{align}
where $\alpha_b$ is a non-dimensional factor, relating the bending and the tangential stiffnesses.

Similarly, the twisting moment vector $M_{ti}$ is described in Equation \ref{Twisting Moment Equation}. An elastic stiffness $K_t=K_{t,b}+K_{t,u}$ is required, formulated Equation \ref{Twisting Stiffness Equation}.
\begin{align}
    M_{ti} &= K_t\Delta\theta_{ti} \label{Twisting Moment Equation}\\
    K_t &= \alpha_t K_s R^1 R^2=2\alpha_t\nu E_m\frac{\left(R^1R^2\right)^2}{R^1+R^2} \label{Twisting Stiffness Equation}
\end{align}
where $\alpha_t$ is a non-dimensional factor, relating the twisting and the tangential stiffnesses. The bending and the twisting resistances aim to reproduce the shape of the grain \cite{Ai2011, Mollon2020, SacMorane:Rolling} as spheres are used in this framework (numerically more efficient).

These contact laws are linear in the elastic domain. Furthermore, a Coulomb friction limit is introduced at the contact level, see Equation \ref{Mohr Coulomb Criteria}

\begin{equation}
    \overline{F_{si}}\leq \mu \; \overline{F_{ni}}
    \label{Mohr Coulomb Criteria}
\end{equation}
where $\mu$ is the friction coefficient between two particles. This friction coefficient is constant before and after the bond breakage.

%The cementation of the sample is modeled as a cohesion between the grains. 
The bond exists until one of the two criteria presented in Equation~\ref{Bond criteria} is not verified.

\begin{align}
    \overline{F_{si}}&\leq \mu \; \overline{F_{ni}} + \sigma_s A_b \text{ (Shear condition)}\nonumber\\
    \overline{F_{ni}}&\leq \sigma_n A_b \text{ (Tensile condition)}
    \label{Bond criteria}
\end{align}
where $\sigma_s$ is the shear strength of the bond, $\sigma_n$ is the tensile strength of the bond and $A_b$ is the surface of the bond. 
It is worth noting that no criterion is employed for the compression condition.
This value of the bond surface $A_b$ is determined initially and is reduced during the simulation, see Section \ref{Numerical model}. During the debonding phenomenon, the mechanical rupture of the bond can be induced by a stress reorganization, increasing the tensile/shear force transmitted in the contact. Furthermore, the rupture can occur even if the tensile/shear force transmitted stays constant as the bond surface $A_b$ decreases during the dissolution. The last rupture case is due to chemistry when the bond surface reaches $A_b=0\,m^2$.
Once the bond breaks (by mechanics or chemistry), the bond surface $A_b$ is set to $0\,m^2$ and the tensile stiffness of the contact is deactivated.

%%=======================================================%%

\section{Numerical model}
\label{Numerical model}

\begin{figure}[ht]
    \centering
    \includegraphics[width=0.95\linewidth]{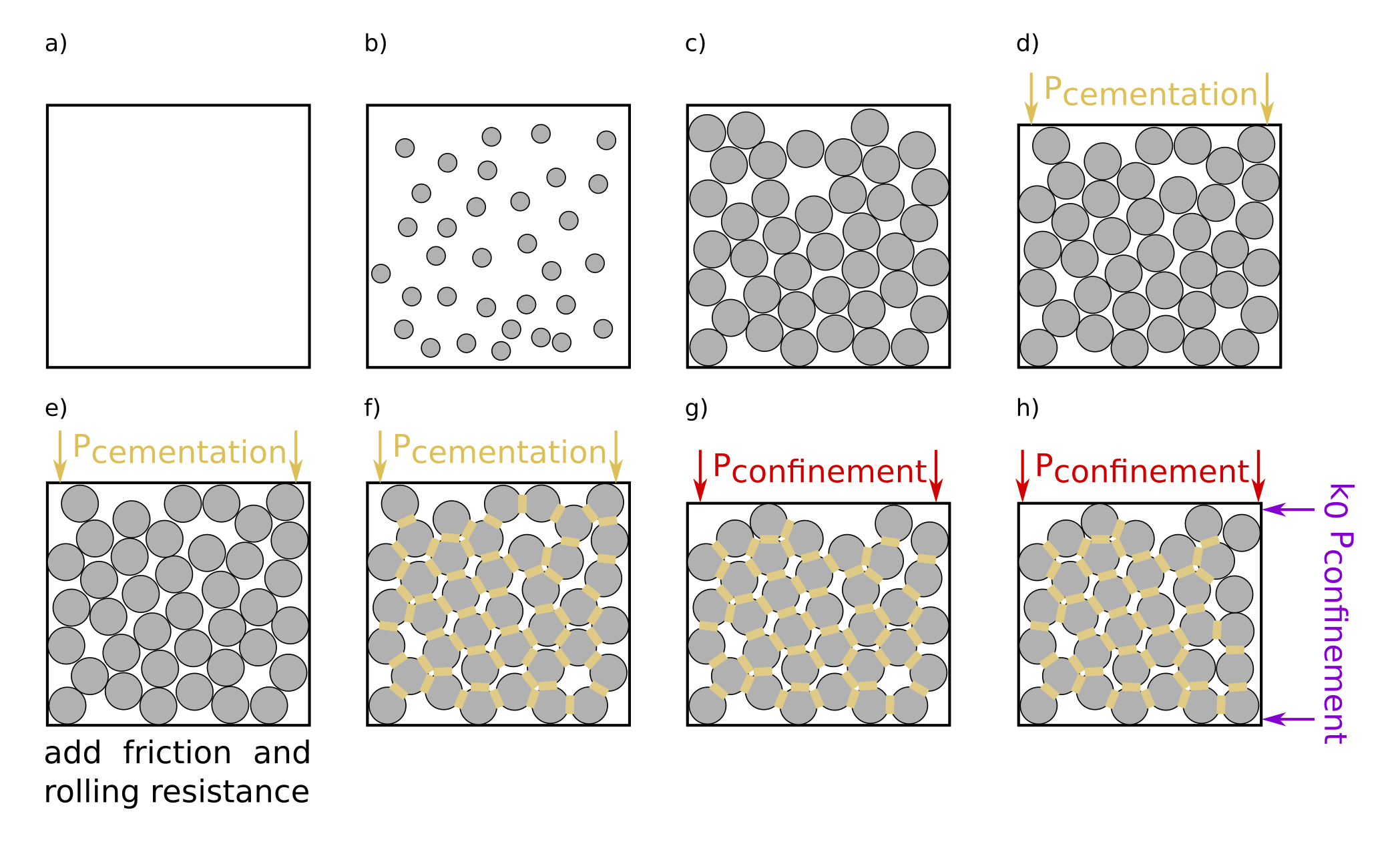}
    \caption{Initial condition algorithm: a) A box is created. b) All the particles are generated with a reduced radius. c) The different radii increase steeply until the final values are reached. d) A vertical pressure $P_{cementation}$ is applied by moving the top plate. e) The friction, the twisting resistance, and the bending resistance are activated. f) The cementation is applied by generating bonds at the contacts. g) A vertical pressure $P_{confinement}$ is applied by moving the top plate. h) An initial stress state $k_0$ is applied by moving the lateral plate.}
    \label{Initial condition algorithm}
\end{figure}

\begin{figure}[ht]
    \centering
    a) \includegraphics[width=0.4\linewidth]{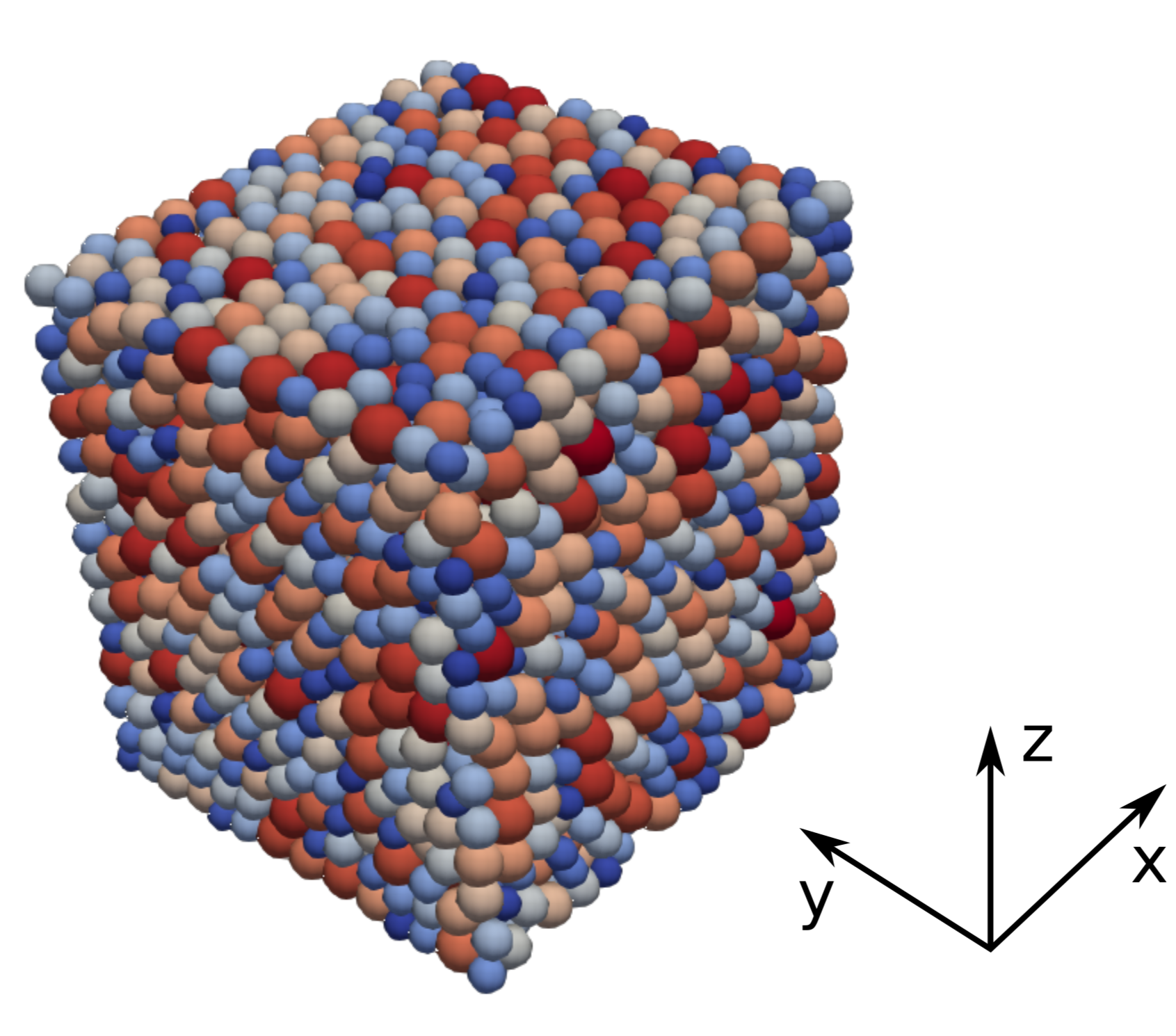} ~
    b) \includegraphics[width=0.4\linewidth]{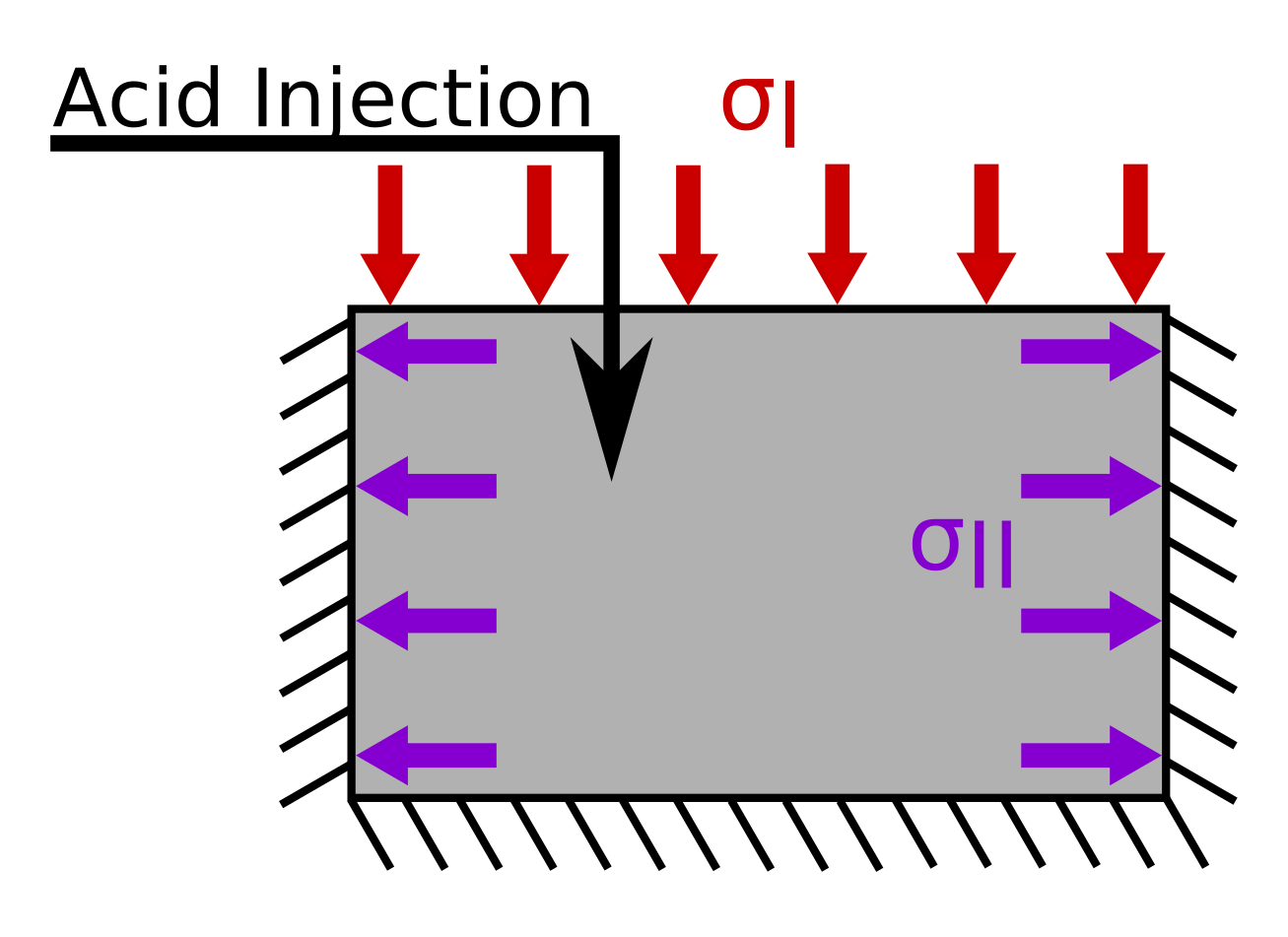}
    \caption{a) An example of an initial configuration obtained by the algorithm presented in Figure~\ref{Initial condition algorithm}. b) The sample is then in oedometric conditions: the lateral and bottom walls are fixed and a constant vertical pressure is applied at the top during the injection of the fluid that induces debonding.}
    \label{Example Configuration}
\end{figure}

The Discrete Element Model is solved using the \emph{YADE} open source software \cite{YADE}.
Some examples of scripts used are available on GitHub: \url{https://github.com/AlexSacMorane/YADE\_oedo\_acid}.
The initial condition algorithm is presented in Figure \ref{Initial condition algorithm}. A $2,8~\times~2,8~\times~2,8$~mm box is generated. Subsequently, $3000$ particles are generated, despite $1600$ would suffice to obtain a representative elementary volume \cite{OSullivan2011,Sarkis2022}. It is crucial to note that the grains are initially incorporated with a radius smaller than the final one. Indeed, a radius expansion algorithm is applied to generate the initial condition \cite{OSullivan2011}. The objective of this algorithm is to verify a uniform grain size distribution similar to \cite{Sarkis2022} ($R_{min}=75$~$\mu$m and $R_{max}=125$~$\mu$m). Once the particles have reached their final dimension, the position of the top wall is controlled in order to apply a vertical pressure equal to $P_{cementation}$ (see Equation \ref{Control Plate} for details). The positions of the remaining walls are maintained at a fixed position. Once the vertical pressure is equal to $P_{cementation}$, the friction between the grains, the twisting resistance, and the bending resistance are activated. The top wall is then controlled to attain once more a vertical pressure equal to $P_{cementation}$, while the other walls remain fixed.
Subsequently, the cementation between the particles is applied. At each contact, a random draw is conducted with a probability equal to $p_c$ to ascertain whether a bond is formed. The bonds are defined by a shear strength $\sigma_s$ and a tensile strength $\sigma_n$. Additionally, the bonds have a surface area $A_b$ obtained by a lognormal distribution presented in Appendix \ref{Lognormal Distribution} and defined by the parameters $m_{log}$ and $s_{log}$ \cite{Sarkis2022}.
The parameters $p_c$, $m_{log}$, and $s_{log}$ characterize the initial degree of cementation, see Table \ref{Parameters Used Cementation}.
Subsequently, the position of the top wall is controlled in order to apply a vertical pressure equal to $P_{confinement}$ (see Equation \ref{Control Plate} for details). The remaining walls are maintained in a fixed position. Subsequently, the position of one lateral wall (the remaining ones are fixed) is controlled to reach an initial coefficient of lateral earth pressure $k_0={\sigma_{II}}/{\sigma_I}$ ($\sigma_I$ is the vertical pressure, $\sigma_{II}$ is the lateral pressure). The sample is subjected to oedometric conditions (the control of the top plate is active and it aims at verifying the vertical confining pressure at any given instant) \cite{SacMorane:PFDEM, Castellanza2004}. Once the lateral pressure has been obtained, the position of the moving lateral plate is fixed to verify the oedometric conditions (fixed lateral walls), and the control of this element is deactivated. 
The initial configuration is reached, and the destabilization due to the debonding phenomenon starts.
An illustrative example of an initial configuration is provided in Figure~\ref{Example Configuration}.

%\newpage

Bond breakage can be triggered during the initialization of the test, preceding the commencement of the bond dissolution through acid injection. The portion of the broken bond is given Table \ref{Bond Breakage Initialization Table} in \ref{Bond Breakage Initialization Section}. Bond breakage will inevitably occur as a consequence of the initialization algorithm, as illustrated in Figure \ref{Initial condition algorithm}. 
The cementation occurs at a specified pressure $P_{cementation}$, and the pressure is then elevated to the confining pressure $P_{confinement}$ before the start of the test. This increase is the source of some bond breakage (even prior to the commencement of dissolution).  
Moreover, the objective is to reproduce numerically the experiments. In an oedometric test with acid injection (bond dissolution), the sample is initially cemented, placed in the oedometer, loaded at confining pressure, and then subjected to acid injection (bond dissolution) \cite{Castellanza2004}. In the experiments, the step of loading to the confining pressure also induces bond breakage.
The occurrence of bond breakage is more prevalent in samples with $k_0^{init}\leq k_0^{attr}$.
Indeed, the value of $k_0$ is approximately equal to $k_0^{attr}$ prior to the formation of the bonds. In order to achieve the objective of $k_0^{init}\leq k_0^{attr}$, it is necessary to extend the lateral dimension of the sample by moving the lateral wall. The combination of this extension with the application of the confining pressure at the top results in shear and tensile conditions at the contact levels. Consequently, bonds breakage occurs if one of the conditions presented in Equation \ref{Bond criteria} is reached.
This phenomenon occurs less frequently in the case of  $k_0^{init}\geq k_0^{attr}$. Indeed, the lateral dimension of the sample is compressed, and it results in less shearing and tensile solicitations at the contact level, but in compression of the bonds. This phenomenon induces less bond breakage since no compression condition is considered as a breakage criterion.

The control of the wall (in the vertical or lateral direction) is achieved through the use of a proportional controller $k_p$ described in Equation \ref{Control Plate}. A maximum speed of the plate $v_{plate}^{max} \left(=0.00005\,R_{mean}/\,dt\right)$ is applied to verify quasi-static conditions \cite{OSullivan2011}, where $R_{mean}$ is the mean radius and $dt$ is the time step defined Equation \ref{Time Step Check}. These aforementioned parameters were determined through a trial-and-error method, minimizing the oscillation of the obtained confining pressure around the target value and maximizing the control kinetics. 

\begin{align}
    v_{plate} &= k_p\times \left(P_{plate}-P_{target}\right)\nonumber\\
    v_{plate} &\leq v_{plate}^{max}
    \label{Control Plate}
\end{align}
where $v_{plate}$ is the velocity of the plate (in the direction of the control), $k_p$ is the value of the proportional controller, $P_{plate}$ is the pressure applied on the plate (in the direction of the control), $P_{target}$ is the targeted pressure applied on the plate (in the direction of the control) and  $v_{plate}^{max}$ is the maximum velocity of the plate.
The walls are considered to have no friction, cohesion, bending resistance, and twisting resistance with the grains. 

\begin{algorithm}[ht]
\caption{The weathering of the rock is modeled as a debonding phenomenon.}
\label{Scheme Dissolution}
\begin{algorithmic}
\While{bonds are remaining} 
    \State Dissolve the bonds \Comment{the bonds' surfaces are decreased} 
    \State Update the mechanical properties, considering the current cementation
    \While{DEM equilibrium not reached} \Comment{granular reorganization occurs} 
        \State Compute solicitations
        \State Solve the momentum balances
        \State Check the equilibrium criteria for DEM
    \EndWhile
    \State Take a snapshot of the configuration \Comment{several parameters are tracked}
    \State Count the number of bonds still existing
\EndWhile
\end{algorithmic}
\end{algorithm}

Once the initial configuration has been established, the dissolution process commences. As presented in Algorithm \ref{Scheme Dissolution}, these steps are subdivided into distinct phases. Initially, the bonds are dissolved, with the surface area of each intact bond reduced (the magnitude of the reduction must be sufficiently minor to ensure representativeness). Subsequently, the reorganization of the grains occurs until an equilibrium state is reached. This mechanical equilibrium is determined by two indices:
\begin{itemize}
    \item the unbalanced force (defined as the ratio of the mean summary force on bodies and mean force magnitude on interactions) is smaller than a criterion value (here $0.01$).
    \item the difference between the pressure applied on the top plate and the targeted pressure is smaller than a criteria value (here $0.01\,P_{target}$). 
\end{itemize}

It is crucial to provide a commentary on the concept of grain reorganization. In the configuration under investigation, the positions of the particles remain largely unchanged. Indeed, the sample is dense and in oedometric conditions. Furthermore, it is assumed that no grain breakage occurs. Nevertheless, some fluctuations in the organization do occur, resulting in an evolution in the chain force structure.

Once the equilibrium is reached, a snapshot of the sample is taken. This enables the tracking of the evolution of the various trackers presented in Section~\ref{Results Section}. These steps are repeated until all the bonds have disappeared. It should be noted that a bond can disappear because of the dissolution (the surface is null or negative) or the mechanical loading (criteria presented in Equation \ref{Bond criteria} are reached).

The parameters used in this paper are presented in Tables \ref{Parameters Used} and \ref{Parameters Used Cementation}. The influence of several parameters such as the vertical confinement pressure $P_{confinement}$ \cite{Hoek2007}, the initial coefficient of lateral earth pressure $k_0^{init}$ \cite{Hoek2007, Taherynia2016, Demir2018} and the degree of cementation \cite{Sarkis2022} are investigated in this paper.

\begin{table}[ht]
    \centering
    \begin{tabular}{|l|p{0.5\linewidth}|}
        \hline
        $P_{confinement}$ & $0.1$ - $1$ - $10$ MPa  \\
        $P_{cementation}$& $0.01 P_{confinement}$\\
        Initial $k_0$& $0.2$ - $0.7$ - $1.5$\\
        \hline
        Cementation & 2T - 2MB - 11BB - 13BT - 13MB\\
        \hline
        Bond dissolution & $0.02\,e^{m_{log}}$ ($0.05\,e^{m_{log}}$ when $95$ \% of the initial number of bonds are broken)\\  
        \hline
    \end{tabular}
    \caption{Presentation of the parameters used (see Table \ref{Parameters Used Cementation} for the details about the cementation).}
    \label{Parameters Used}
\end{table}

\begin{table}[ht]
    \centering
    \begin{tabular}{|p{0.15\linewidth}|c|c|c|c|c|c|}
        \multicolumn{2}{c}{}&\multicolumn{2}{|p{0.13\linewidth}|}{Lightly cemented}&\multicolumn{1}{p{0.13\linewidth}}{Medium cemented}&\multicolumn{2}{|p{0.13\linewidth}|}{Highly cemented}\\
        \hline
        Sample&Untreated&2T&2MB&11BB&13BT&13MB\\
        \hline
        Density (kg/m$^3$)&\multicolumn{6}{c|}{$2650$}\\
        \hline
        $E_m$ (MPa)&$80$&$300$&$320$&$760$&$860$&$1000$\\
        \hline
        $\nu$ (-)&\multicolumn{6}{c|}{$0.25$}\\
        \hline
        $\alpha_b$ (-)&\multicolumn{6}{c|}{$0.5$}\\
        \hline
        $\alpha_t$ (-)&\multicolumn{6}{c|}{$0.5$}\\
        \hline
        $\mu$ (-)&\multicolumn{6}{c|}{$0.36$}\\        
        \hline
        $p_c$ (\%)&$0$&$13$&$88$&$98$&$100$&$100$\\  
        \hline
        $m_{log}$ (-)&$0$&$6.79$&$7.69$&$8.01$&$8.44$&$8.77$\\
        \hline
        $s_{log}$ (-)&$0$&$0.70$&$0.60$&$0.88$&$0.92$&$0.73$\\   
        \hline
        $\sigma_s$ (MPa)&\multicolumn{6}{c|}{$6.6$}\\ 
        \hline
        $\sigma_n$ (MPa)&\multicolumn{6}{c|}{$2.75$}\\ 
        \hline
    \end{tabular}
    \caption{Mechanical parameters depending on the cementation (extracted from \cite{Sarkis2022}).}
    \label{Parameters Used Cementation}
\end{table}

\begin{figure}[h]
    \centering
    \includegraphics[width=0.6\linewidth]{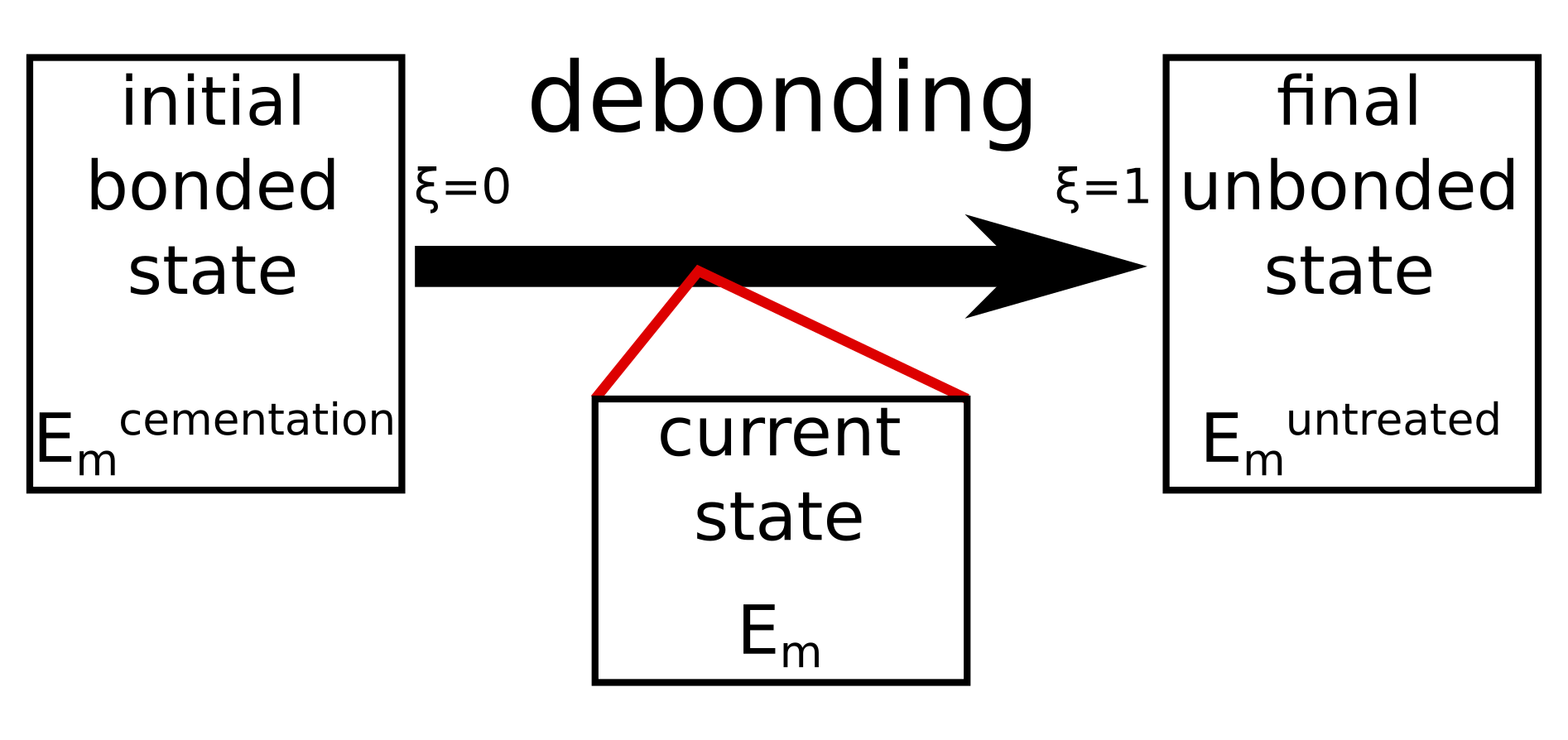}
    \caption{During the debonding phenomenon, the current state is interpolated from the initial (cemented) and final (unbonded) states.}
    \label{Debonding Interpolation}
\end{figure}

To explore the effect of the cementation, five distinct initial cementations, exhibiting disparate microscopic parameters, are considered. During the debonding process, the bond surfaces $A_b$ are reduced, and the sample aims to reach the untreated granular state. The reduction of the bond surfaces results in a reduction in the strength of the cemented contacts, see Equation \ref{Bond criteria}. Such weakening may ultimately result in the mechanical rupture of the bond. Moreover, in the event that the bond remains intact and its surface area reaches $A_b=0m^2$, it may also undergo a chemical rupture. 
These ruptures weaken the sample, and the current state during the debonding phenomenon is interpolated from the only known states: the initial cemented state and the final unbonded state, see Figure \ref{Debonding Interpolation}.

Table \ref{Parameters Used Cementation} highlights the fact that the Young modulus $E_m$ exhibits variation between the unbonded and bonded states. To ensure the sample is representative, and because of the data available in \cite{Sarkis2022}, an interpolation between the two known states is performed to characterize the current state. This approach is commonly applied in the literature to model chemical damage at the macroscale \cite{Gajo2015,Gajo2019,Xang2023,Viswanath2020b}.
In particular, the Young modulus (and thus, the different contact springs $K_n$, $K_s$, $K_r$, $K_t$) is updated following Equation \ref{Young Reduction}. This assumption results in the dependency of the contact scale on the sample scale through the debonding factor $\xi$. In this investigation, a linear approximation is employed as it represents the most straightforward relation available for a system comprising only two known states. Should supplementary data become available, this relation can be readily modified in accordance with the new information.
The impact of this stiffness reduction will be examined in this paper in Section \ref{Young modulus Reduction Results} and a discussion will be presented in Section \ref{Discussion Young modulus reduction}.

\begin{equation}
    E_m = (E_m^{cementation}-E_m^{untreated})\times (1-\xi) + E_m^{untreated}
    \label{Young Reduction}
\end{equation}
where $E_m$ is the current Young modulus used in the simulation, $E_m^{cementation}$ is the initial Young modulus depending on the initial degree of cementation, $E_m^{untreated}$ is the Young modulus without cementation (here $80$ MPa), and $\xi$ is the debonding factor defined as the ratio of the bonds dissolved (by dissolution or by loading) over the initial number of the bonds. As the time step employed for the temporal integration depends on the Young modulus, see Equation \ref{Time Step Check}, it becomes important to update the time increment $dt$ with the dissolution of the bonds.
The relationship between the contact scale ($E_m$) and the sample scale ($\xi$) could be a matter of contention. An alternative approach would have been to consider a local surface stiffness $E_m(A_b)$ at each contact, with the stiffness dependent on the bond surface \cite{Sun2016,Sun2018}. This approach is explored in Section \ref{Young from Ab}. 
However, this relationship is not available in the data set used \cite{Sarkis2022}, and can only be extrapolated. Consequently, the model choice has been to interpolate the current state, see Figure \ref{Debonding Interpolation}. Nevertheless, contact strength depends on the local bond surface area $A_b$, see Equation \ref{Bond criteria}.

%%=======================================================%%

\section{Results}
\label{Results Section}

The results of all the simulations performed (plus some complementary investigations) are presented in the following Subsections. Two main variables are investigated:
\begin{itemize}
    \item the coefficient of lateral earth pressure $k_0=\sigma_{II}/\sigma_I$, where $\sigma_{I}$ is the vertical stress and $\sigma_{II}$ is the lateral stress. This parameter is a proxy of the state of stress inside the sample \cite{Santamarina2009, Santamarina2014}.   
   \item the debonding variable $\xi=1-n_{bond}/n_{bond}^0$, where $n_{bond}$ is the number of the bond and $n_{bond}^0$ is the number of the bond after the initial configuration. This variable is equal to $0$ when the simulation starts (no bond broken) and equal to $1$ when the simulation ends (all bonds broken).
\end{itemize}
Thanks to the DEM, much more data are available to understand the mechanisms operating at the microstructural level such as the coordination number, the mode of rupture of the bonds or the mean force transmitted in the contact, among others. These data will be used if necessary to understand our results. 

%%===========================%%

\subsection{The existence of an attractor value $k_0^{attr}$}

\begin{figure}[ht]
    \centering
    a) \includegraphics[width=0.45\textwidth]{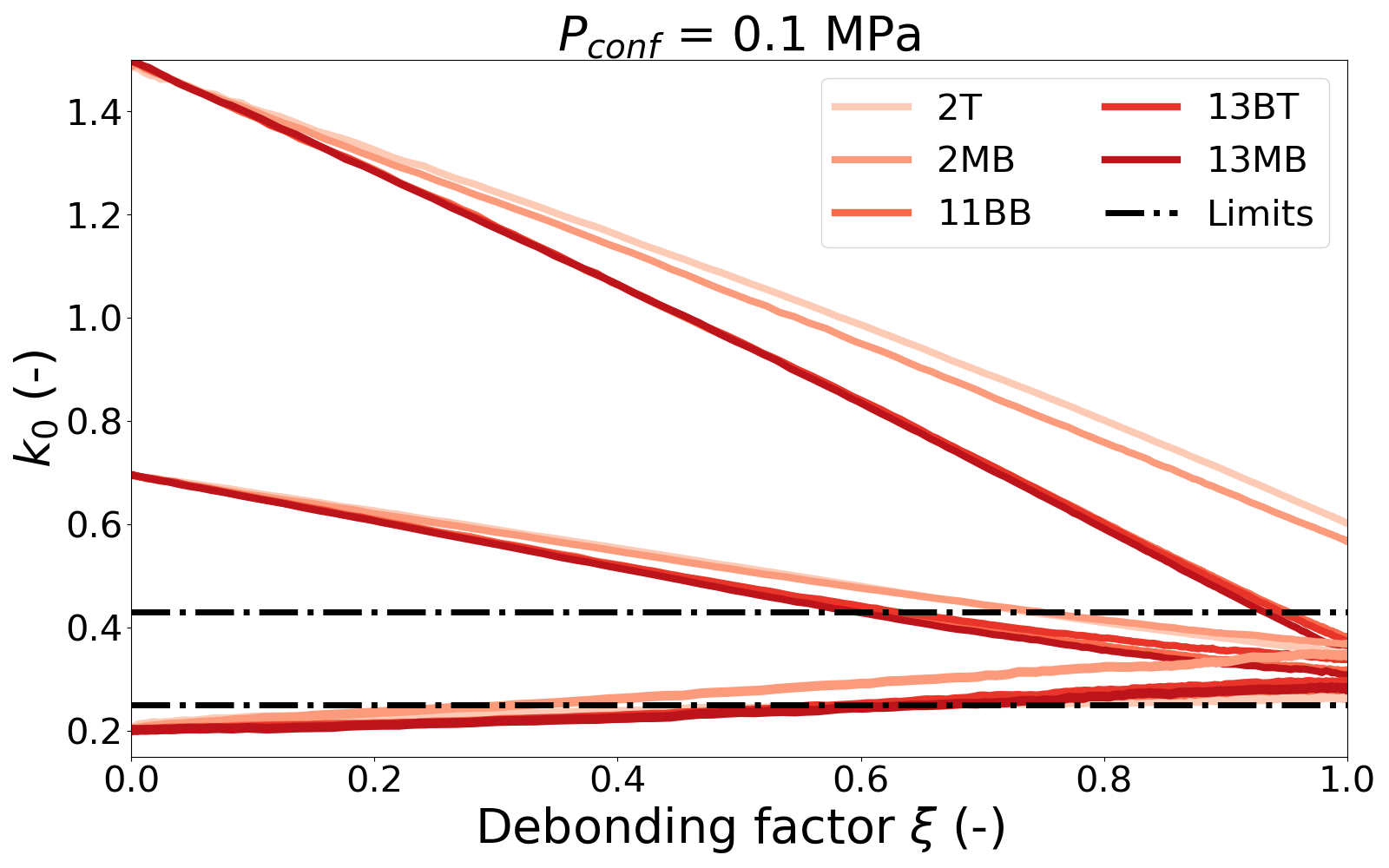}
    b) \includegraphics[width=0.45\textwidth]{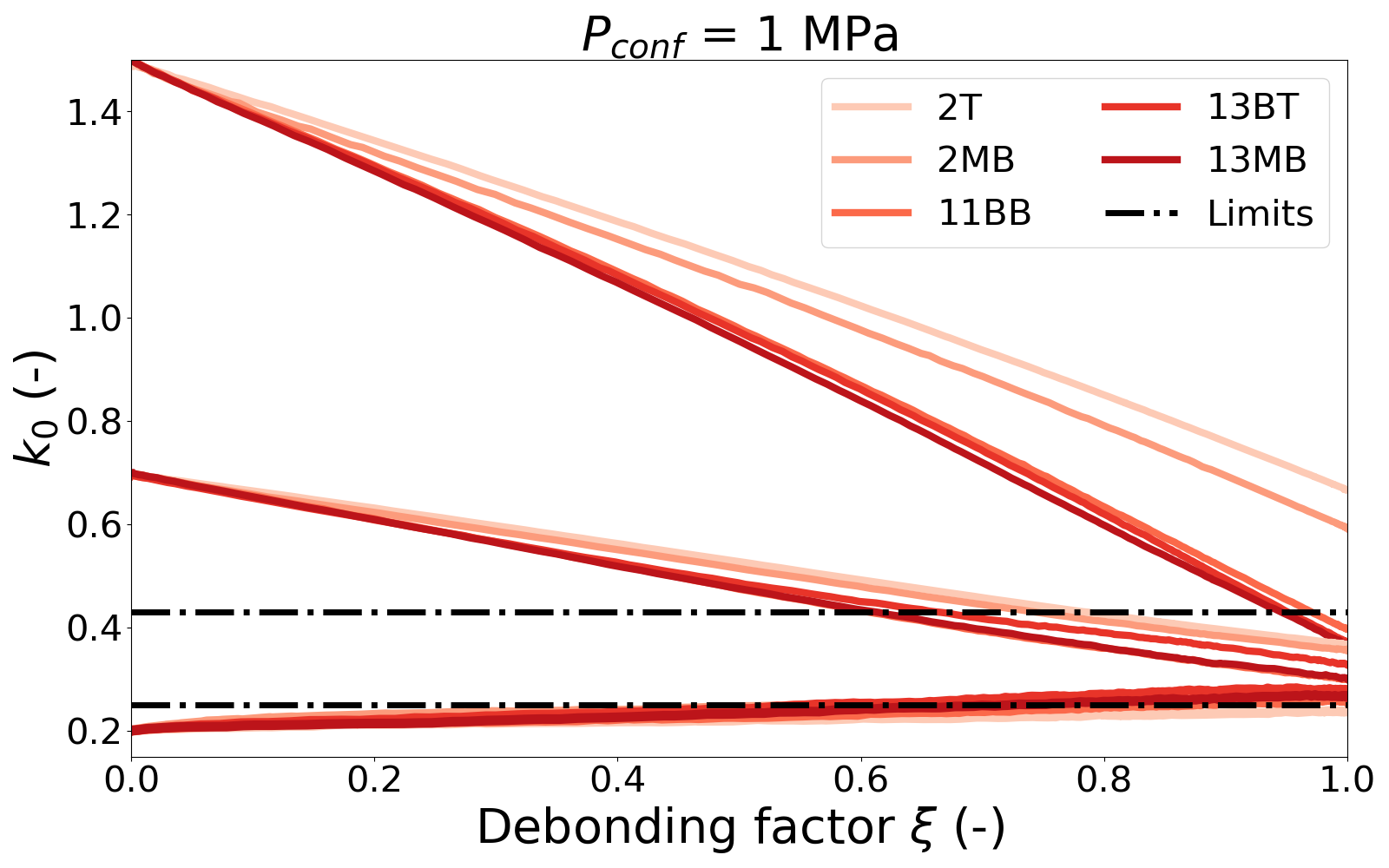}\\
    c) \includegraphics[width=0.45\textwidth]{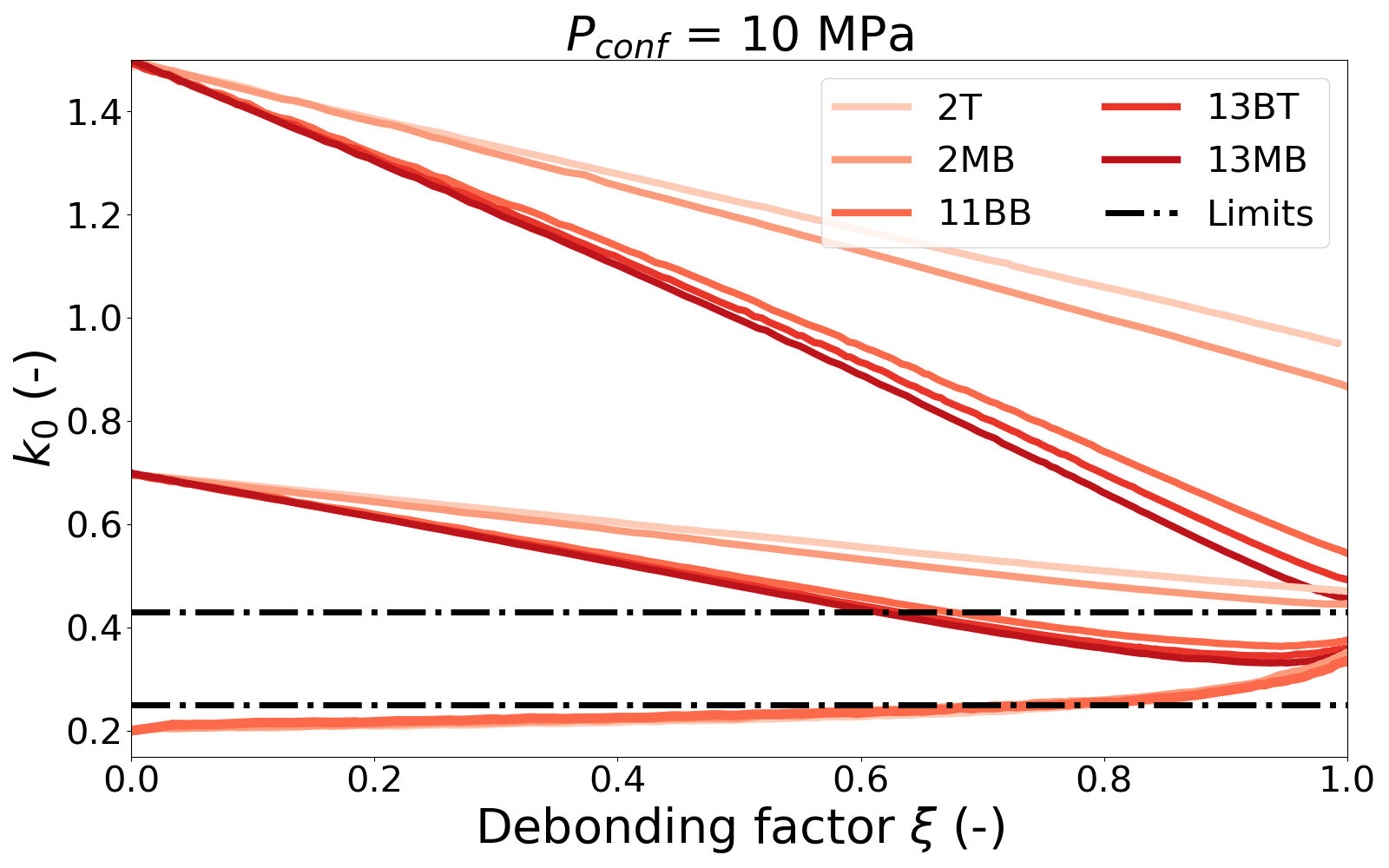}
    \caption{Evolution of $k_0$ with the debonding factor $\xi$ for $P_{confinement}=$  a) $0.1$, b) $1$ and c) $10$ MPa. Limits considering $k_0^{attr}=\nu/(1-\nu)$ are plotted with $\nu=0.2-0.3$.}
    \label{k0 attractor Figure}
\end{figure}

The evolutions of $k_0$ are gathered in Figure \ref{k0 attractor Figure} for the five initial degrees of cementation, the three values of $k_0^{init}$ and the three distinct confining pressures $P_{confinement}$.

It appears the $k_0$ aims to reach an attractor value $k_0^{attr}$ with the dissolution of the bonds.
Applying the principle of superposition, it has been shown analytically by \cite{Vaughan1984} that $k_0$ approaches a limit given by $k_0^{attr}=\nu/(1-\nu)$ during weathering and stiffness reduction. Considering a window of $[0.2-0.3]$ for the value of $\nu$ \cite{Cheng2017,Ziccarelli2024,Mortazavi2024}, the attractor value window is $[0.25-0.43]$, in agreement with the results obtained. This evolution of the $k_0$ illustrates the granular reorganization \cite{Santamarina2009, Santamarina2014}. The sample being dense and under oedometric conditions, the positions of the particles remain mostly unchanged. Nevertheless, these small fluctuations affect widely the distribution of forces between particles and thus the stress state of the sample.

\vskip\baselineskip

\begin{figure}[ht]
    \centering
    \includegraphics[width=0.9\textwidth]{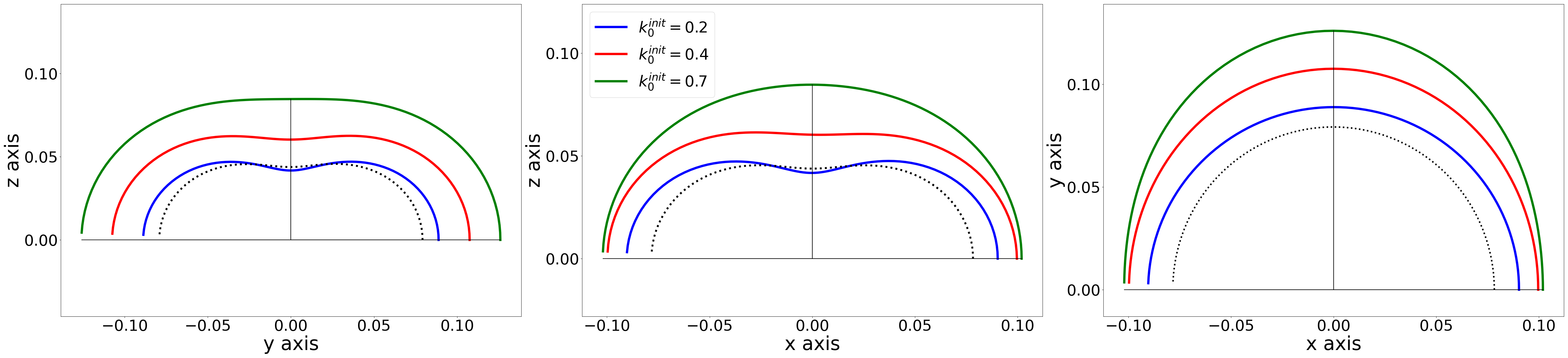}\\
    \caption{Distribution function $f^n(n_i)$ at initial and attractor conditions (dotted). This function represents the average normal force on a specific direction.}
    \label{Distribution Function Normal Figure}
\end{figure}

To understand better the attractor configuration, the fabric description of the sample and its evolution are investigated. Three simulations are considered: a 13MB sample with a confining pressure $P_{confinement}=1$ MPa and initial values of $k_0= 0.2$, $0.4$ or $0.7$.
While contact-based fabric tensors are frequently referenced in the literature \cite{Radjai1998, Shi2018, Liu2020}, this study employs a normal forces-based fabric tensor $F^n_{ij}$ \cite{Liu2020, Guo2013, Gong2019}. Indeed, the normal forces-based fabric tensor underlines information on the fabric anisotropy, weighted by the norm of the normal force transmitted at the contacts. Being dense and under oedometric conditions, the normal forces are dominant in the stress transmission. The debonding process results in an evolution of chain forces, even if the position of the grains remains similar. The primary cause of this evolution can be attributed to the modification of normal forces. To quantify this variation, a normal forces-based tensor is defined as the average of the outer product of the normal forces at contacts, see Equation \ref{Normal Fabric Tensor Equation}.

\begin{equation}
    F^n_{ij} = \frac{1}{N_c}\sum\limits_c \frac{f^n n_i n_j}{1 + D_{kl} n_k n_l}
    \label{Normal Fabric Tensor Equation}
\end{equation}
where $N_c$ is the total number of contacts, $c$ is a contact in the sample, $f^n$ is the norm of the normal force of the contact $c$, $n_i$ is the unit vector of the contact direction and $D_{kl}$ is the fabric tensor of the second order defined in Equation \ref{2nd Order Fabric Tensor Equation} \cite{Liu2020}.

\begin{equation}
    D_{ij} = \frac{15}{2}\left(\Phi_{ij}-\frac{1}{3}\delta_{ij}\right)
    \label{2nd Order Fabric Tensor Equation}
\end{equation}
where $\Phi_{ij}$ is the contact-based fabric tensor, defined in Equation \ref{Contact Fabric Tensor Equation} \cite{Radjai1998, Liu2020, Gong2019} and $\delta_{ij}$ is the Kronecker tensor.

\begin{equation}
    \Phi_{ij} = \frac{1}{N_c}\sum\limits_c n_i n_j
    \label{Contact Fabric Tensor Equation}
\end{equation}
where $N_c$ is the total number of contacts, $c$ is a contact in the sample and $n_i$ is the unit vector of the contact direction. A distribution function $f^n(n_i)$ can be defined in Equation \ref{Distribution Function Normal} \cite{Liu2020} and illustrated in Figure \ref{Distribution Function Normal Figure} at initial conditions for $k_0^{init}=0.2$ (blue), $0.4$ (red) and $0.7$ (green) and at the attractor conditions (dotted). Even if the initial value of the $k_0$ influences the probability distribution function at the initial state, it appears that the fabric aims to reach an attractor configuration thanks to the granular reorganization. This attractor configuration is computed considering the average of the three final configurations (individually similar to the attractor).

\begin{equation}
    f^n(n_i) = f_0\left(1+a^n_{ij} n_i n_j\right)
    \label{Distribution Function Normal}
\end{equation}
where $f_0=tr(F^n_{ij})$ and $a^n_{ij}=15/2\left(F^n_{ij}/f_0-\delta_{ij}\right)$. This function represents the average normal force in a specific direction \cite{Guo2013}.

\begin{figure}[ht]
    \centering
    \includegraphics[width=0.9\textwidth]{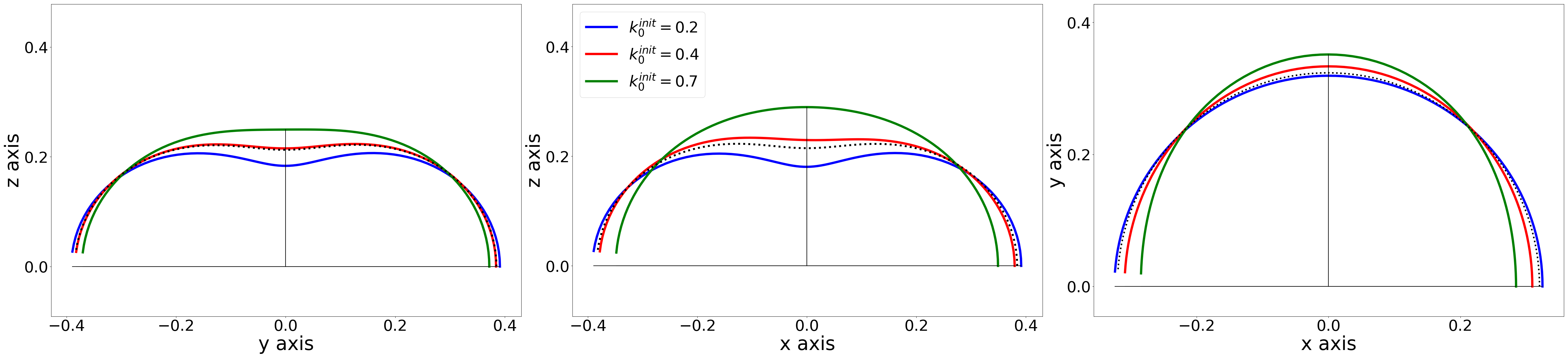}\\
    \caption{Distribution function $\overline{f^n}(n_i)$ at initial and attractor conditions (dotted). This function represents the probability of a contact to exist in a direction weighted by the average normal force in this direction.}
    \label{Normalized Distribution Function Normal Figure}
\end{figure}

It can be observed that the function $k_0^{init}=0.2$ (depicted in blue) is situated within the function $k_0^{init}=0.4$ (depicted in red), which in turn is situated within the function $k_0^{init}=0.7$ (depicted in green). As illustrated in Figure \ref{Mean Force Figure}, the mean force transmitted in contact is found to be larger for the case $k_0^{init}=0.7$ than for either $k_0^{init}=0.4$ or $k_0^{init}=0.2$. This is consistent with the definition of the function $f^n$, representing the average normal force in a specific direction. Furthermore, it is observed that as $k_0^{init}$ increases, the distribution of $f^n$ for the planes yz and xz approaches a circle. The sphericity of $f^n$ indicates that the distribution of the forces is more isotropic, which is consistent with the values of $k_0^{init}$  closer to $1$. This tendency is even clearer in Figure \ref{Normalized Distribution Function Normal Figure}, where the distribution function is normalized.

One can notice the fact that an anisotropy exists on plane xy. Indeed, the initial configuration algorithm consists of moving only one wall to apply $k_0^{init}$, favoring one loading direction, see Figure \ref{Initial condition algorithm}. However, this anisotropy remains insignificant.

\vskip\baselineskip

Then, this probability distribution function is normalized to compare the different organizations. A normalized function $\overline{f^n}$ is defined in Equation \ref{Normalized Distribution Function Normal} and shown in Figure \ref{Normalized Distribution Function Normal Figure}.

\begin{equation}
    \overline{f^n}(n_i) = f_0\frac{1+a^n_{ij} n_i n_j}{\int f^n\,(n_i)\,d\theta}
    \label{Normalized Distribution Function Normal}
\end{equation}
where $\int f^n\,(n_i)\,d\theta$ is the integral of $f_n$ along the plane considered (xy, xz or yz). Thanks to this definition, the perimeter of $\overline{f_n}$ equals $1$ in the plane. This function represents the probability of a contact to exist in the direction $n_i$ weighted by the average normal force in this direction.

The normalization enables us to compare the cases. Looking at planes yz and xz, it is clear that the $k_0^{init}=0.2$ (blue) case must create isotropy ($k_0$ will increase), whereas the $k_0^{init}=0.4$ (red) or $=0.7$ (green) cases must create anisotropy ($k_0$ will decrease). 
As a reminder, an isotropic state of stress is defined as $k_0 = 1$, and it is depicted as a perfect circle.
The same observations can be done on the plane xy, one direction is favored because of the initial condition algorithm, see Figure \ref{Initial condition algorithm}. However, this anisotropy remains insignificant.

\vskip\baselineskip

\begin{figure}[ht]
    \centering
    \includegraphics[width=0.9\textwidth]{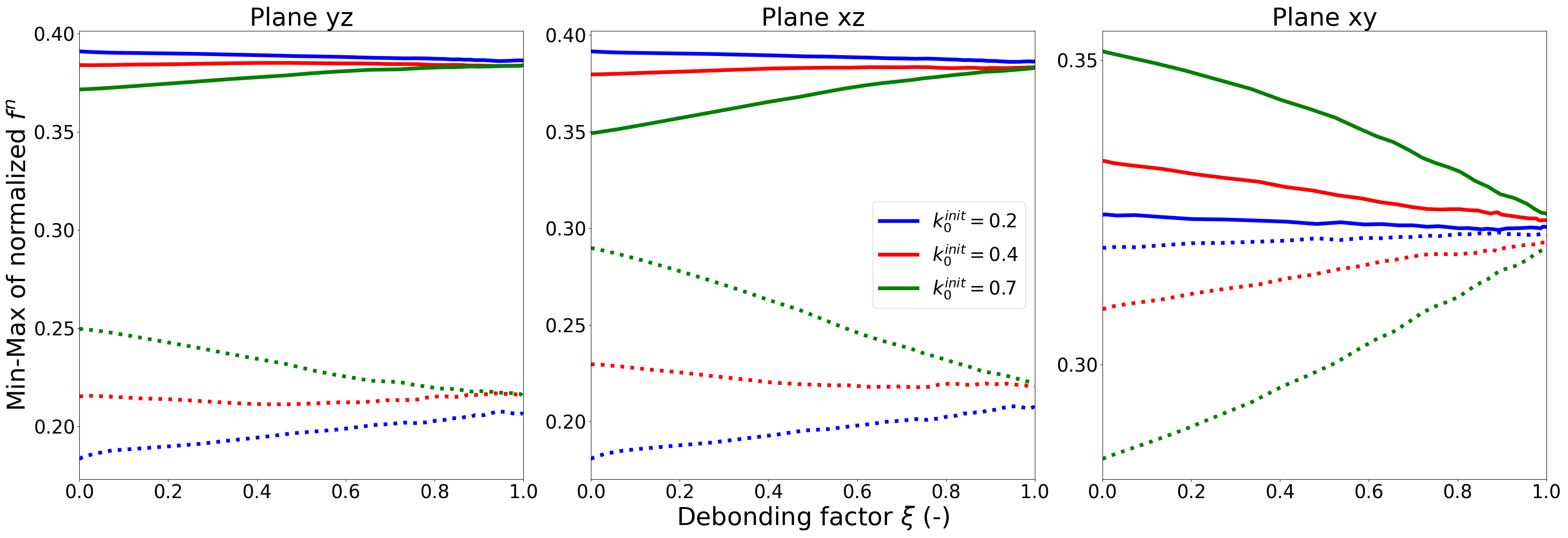}
    \caption{Evolution of the maximum and the minimum values of the normalized probability function $\overline{f^n}(n_i)$ with the debonding factor $\xi$.}
    \label{Max-Min Distribution Function Normal Figure}
\end{figure}

The evolution during the debonding of the shape of the normalized probability function $\overline{f^n}(n_i)$ can be captured by the evolution of the maximum and the minimum values, illustrated in Figure \ref{Max-Min Distribution Function Normal Figure}. 

It appears that in planes yz and xz, the maximum values increase and the minimum values decrease for cases $k_0^{init} =0.4$ and $=0.7$. This distancing of the extrema reveals the creation of anisotropy. On the opposite, the maximum values decrease and the minimum values increase for the case $k_0^{init} =0.2$, revealing the creation of isotropy (closing of the extrema). These observations are in agreement with the evolution of the $k_0$, proxy of the isotropy/anisotropy of the sample \cite{Santamarina2009, Santamarina2014}. 
In plane xy, the extrema are closing for the three cases (creation of isotropy).
The initial anisotropy is generated by the initial configuration algorithm, see Figure \ref{Initial condition algorithm}, and it is diminished with the debonding phenomenon.

\vskip\baselineskip

It is crucial to acknowledge that $k_0$ is computed in this study from the stress applied on a single lateral wall. One might inquire as to how the stress states evolve in the minor and intermediate principal directions. 
Figures \ref{Distribution Function Normal Figure}, \ref{Normalized Distribution Function Normal Figure} and \ref{Max-Min Distribution Function Normal Figure} show that anisotropy exists initially on the plane xy (especially for $k_0^{init} = 0.4$ and $0.7$). This is because of the initialization process, see Figure \ref{Initial condition algorithm}. Indeed, only a single wall (with the normal in the x-direction) is moving to control $k_0$. Minor and intermediate stresses are different, even if they stay similar. Minor stress is in the direction used to compute $k_0$.
With the debonding the anisotropy on this plane disappears, see the attractor line (dotted), and minor and intermediate stresses aim to reach the same value. 

%%===========================%%

\subsection{The existence of two $k_0$ evolution mechanisms}
\label{Young modulus Reduction Results}

\begin{figure}[ht]
    \centering
    a) \includegraphics[width=0.45\textwidth]{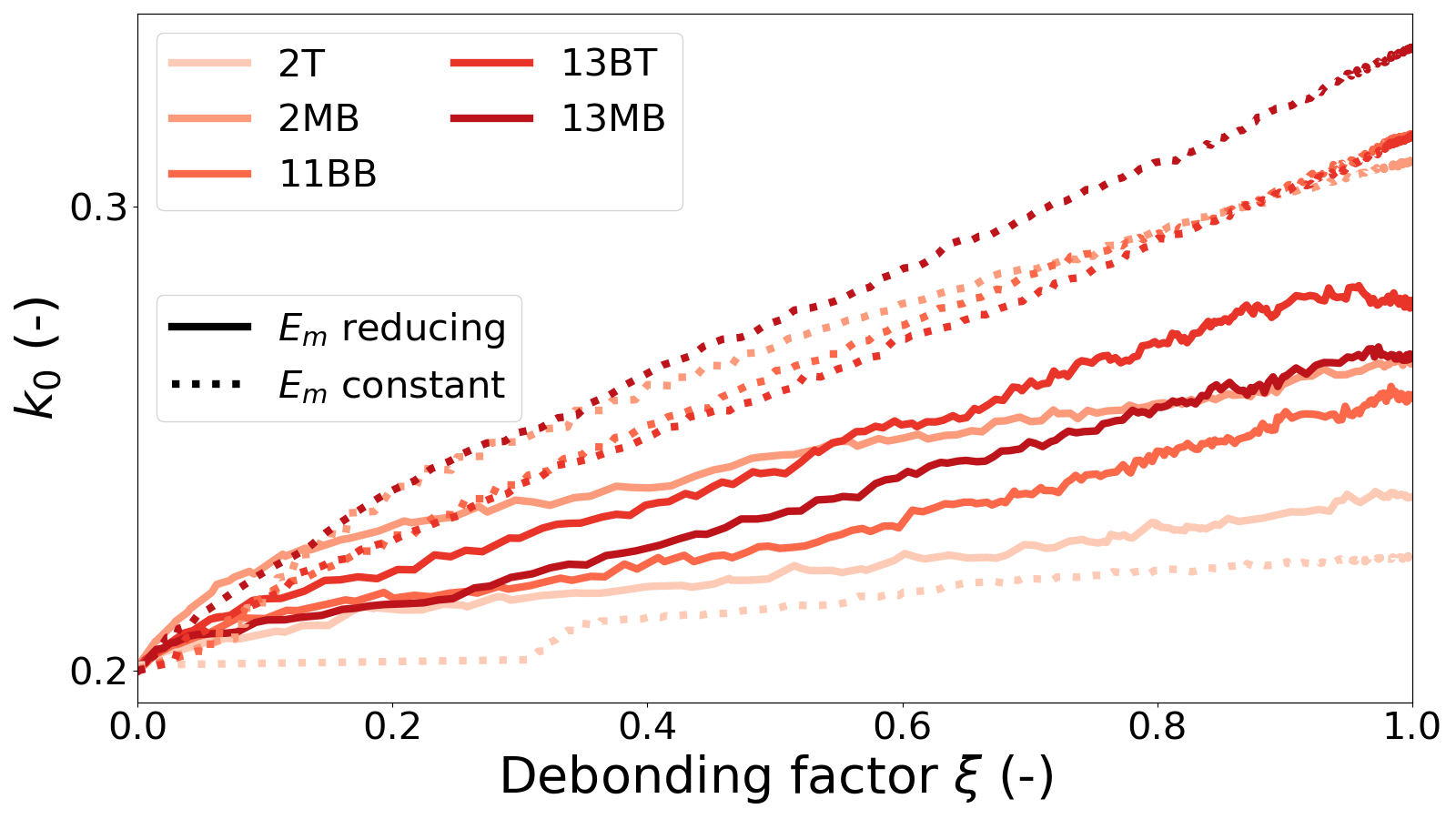}~ 
    b) \includegraphics[width=0.45\textwidth]{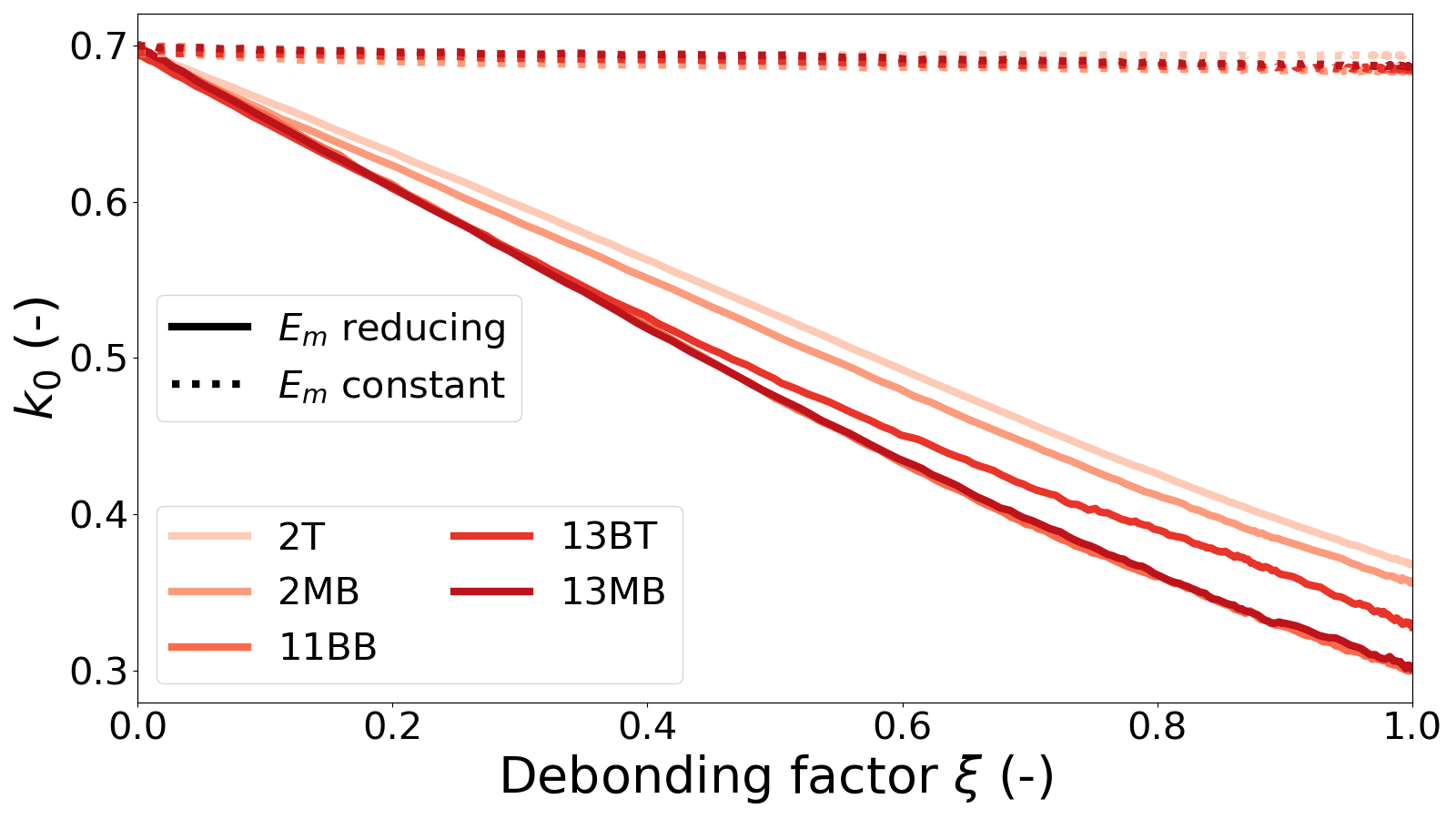}
    \caption{Effect of the Young modulus reduction assumption for different cementations at a confinement pressure $P_{confinement}=1$ MPa with an initial $k_0=$~a)~$0.2$ or b)~$0.7$.} 
    \label{k0 evolution vs Young Modulus Assumption}
\end{figure}

This Section addresses the issue of the reduction in Young modulus during the dissolution process, see Equation \ref{Young Reduction}. It also investigates the mechanisms of the $k_0$ evolution.
The identical simulations conducted in the preceding Section are now executed without the assumption of a reduction in the Young modulus (the Young modulus is held constant throughout the debonding process). The results are presented in Figure \ref{k0 evolution vs Young Modulus Assumption}.

In the case $k_0^{init} \leq k_0^{attr}$, the data illustrated in Figure \ref{k0 evolution vs Young Modulus Assumption}a appears to demonstrate that a grain reorganization occurs ($k_0$ evolves) even when the Young modulus remains constant. On the other hand, Figure \ref{k0 evolution vs Young Modulus Assumption}b illustrates that grain reorganization does not occur ($k_0$ remains constant) when the Young modulus remains constant for $k_0^{init} \geq k_0^{attr}$. In this configuration, the reorganization of the grains is a consequence of the reduction in the material's Young modulus. Two distinct mechanisms appear to emerge, contingent on the state of stress.

\begin{figure}[ht]
    \centering
    \includegraphics[width=0.6\textwidth]{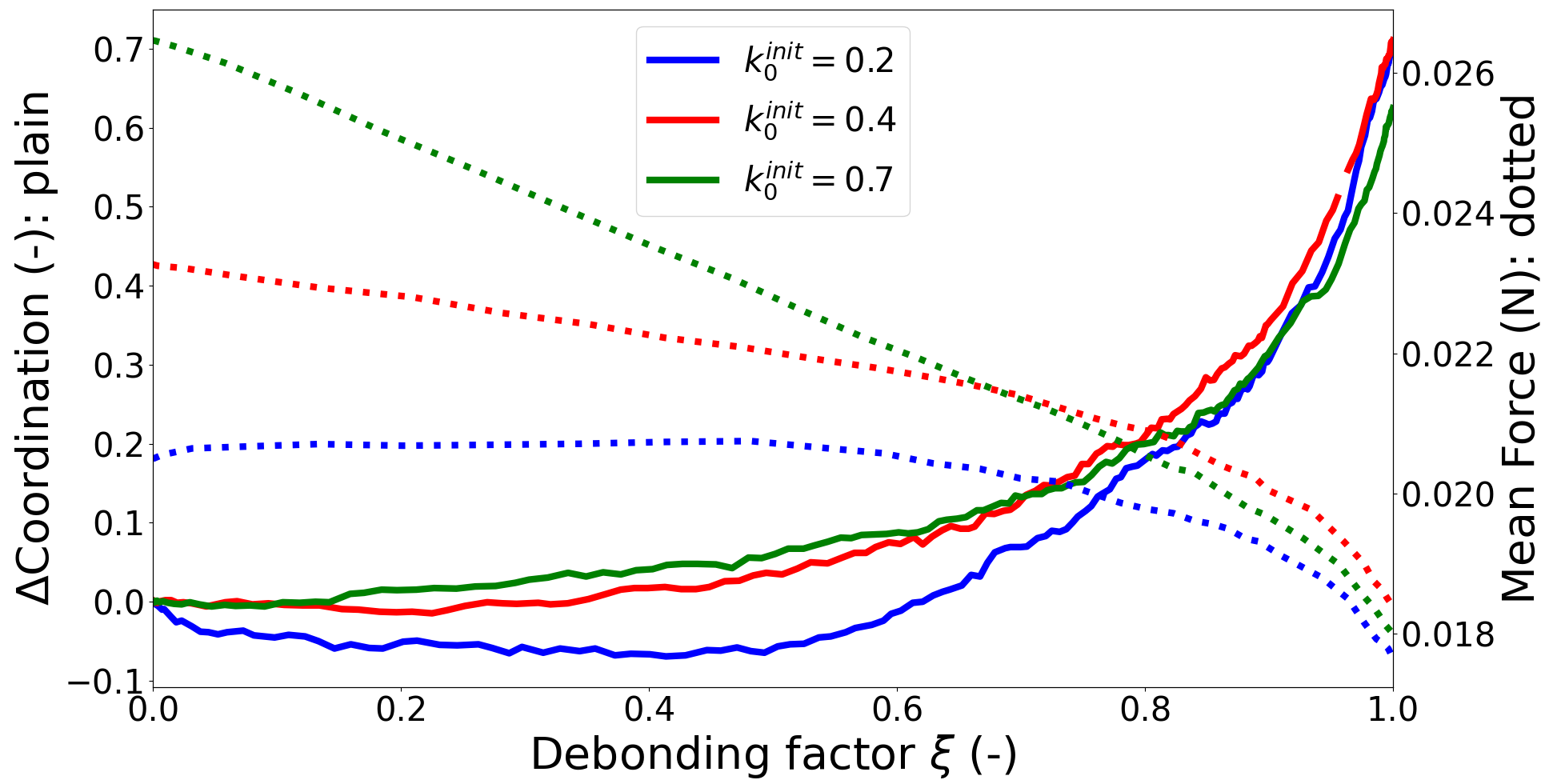}
    \caption{Evolution of the coordination number (plain line), mean number of contacts per grain, with the debonding phenomena. The evolution of the mean force (normal + tangential) transmitted in contacts of the fabric (dotted lines) is plotted in parallel. Notice the coordination number evolutions have been set to start at $0$ to be compared.}
    \label{Mean Force Figure}
\end{figure}

This phenomenon is also highlighted by considering the evolution of the mean force and the coordination number (proxy of the grain organization). Three simulations are considered: a 13MB sample with a confining pressure $P_{confinement}=1$ MPa and initial values of $k_0= 0.2$, $0.4$ or $0.7$. The results are shown in Figure \ref{Mean Force Figure}.

The distinction between the two mechanisms is underscored by the observation that the mean force evolution corresponding to $k_0^{init} = 0.2$ exhibits a distinct trajectory compared to the mean force evolutions associated with $k_0^{init}=0.4$ or $=0.7$.
The first one remains constant ($\approx 0.02$ N), until a threshold value for the debonding is reached ($\xi\approx 0.6$). In contrast, the mean forces associated with $k_0^{init}=0.4$ or $=0.7$ decrease in a linear fashion from the outset of the simulation. A second evolution, exhibiting a more pronounced and less linear trend, can be observed following the attainment of a debonding threshold ($\xi\approx 0.8$). Similarly, it seems that the coordination number corresponding to $k_0^{init}=0.2$ reduces at the outset of the simulation. Whereas, the coordination numbers corresponding to $k_0^{init}=0.4$ and $=0.7$ are intended to increase. This observation served to reinforce the hypothesis that two distinct mechanisms are at play, contingent on the state of stress.

%%===========================%%

\subsection{Influence of the initial stress state on the main grain reorganization mechanism}
\label{Grain Reorganization Mechanism Section}

As emphasized in Section \ref{Young modulus Reduction Results}, the principal grain reorganization mechanism seems to depend on the state of the stress. This reorganization induces an evolution of $k_0$, related to the evolution of the chain force \cite{Santamarina2009, Santamarina2014}. It should be noted that the positions of the particles remain largely unaltered as the sample is dense in oedometric conditions. Nevertheless, the minor fluctuations impact the state of the stress within the sample.

\begin{figure}[h]
    \centering
    \includegraphics[width=0.55\linewidth]{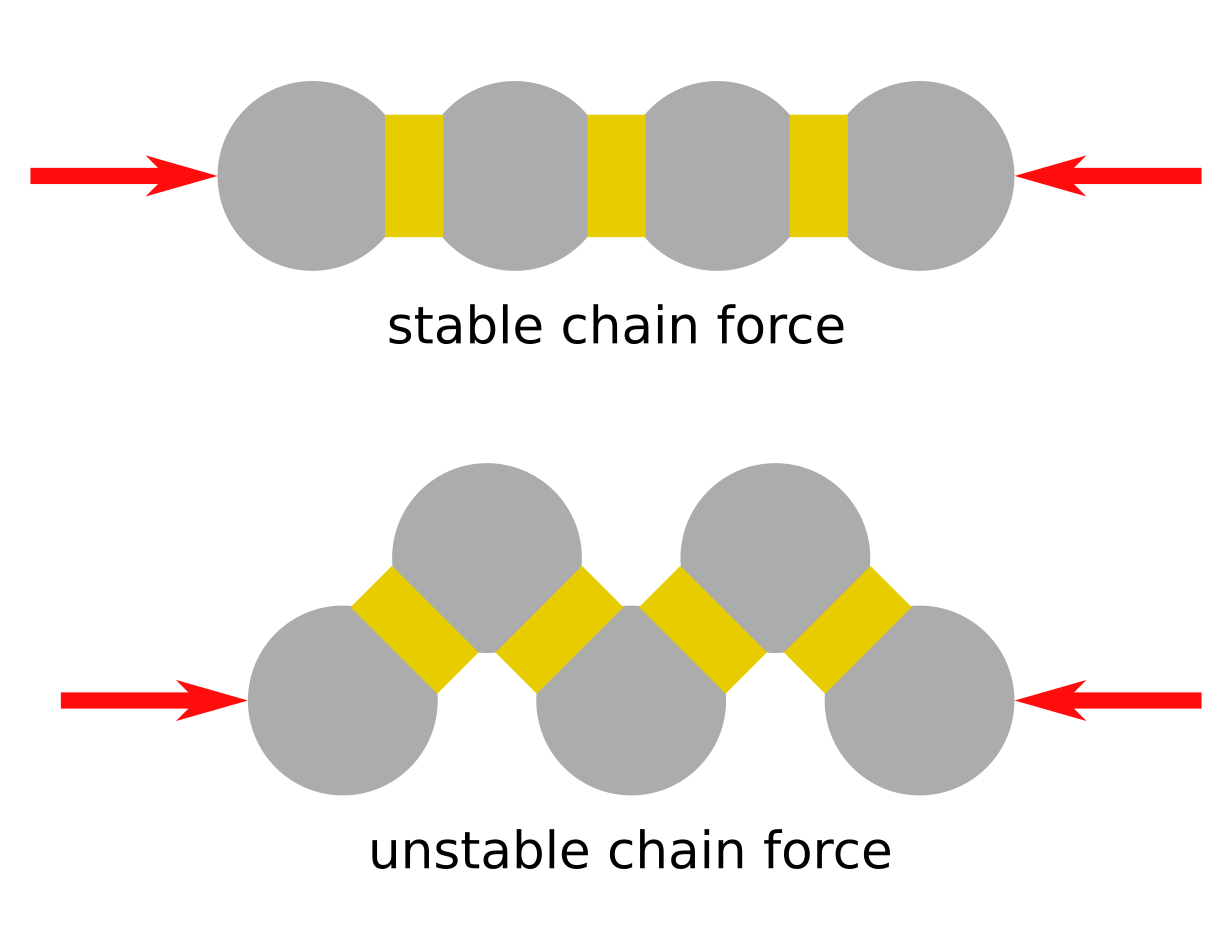}
    \caption{The granular media comprises unstable and stable chain forces.}
    \label{Stable Unstable CF}
\end{figure}

\begin{figure}[ht]
    \centering
    \includegraphics[width=0.8\textwidth]{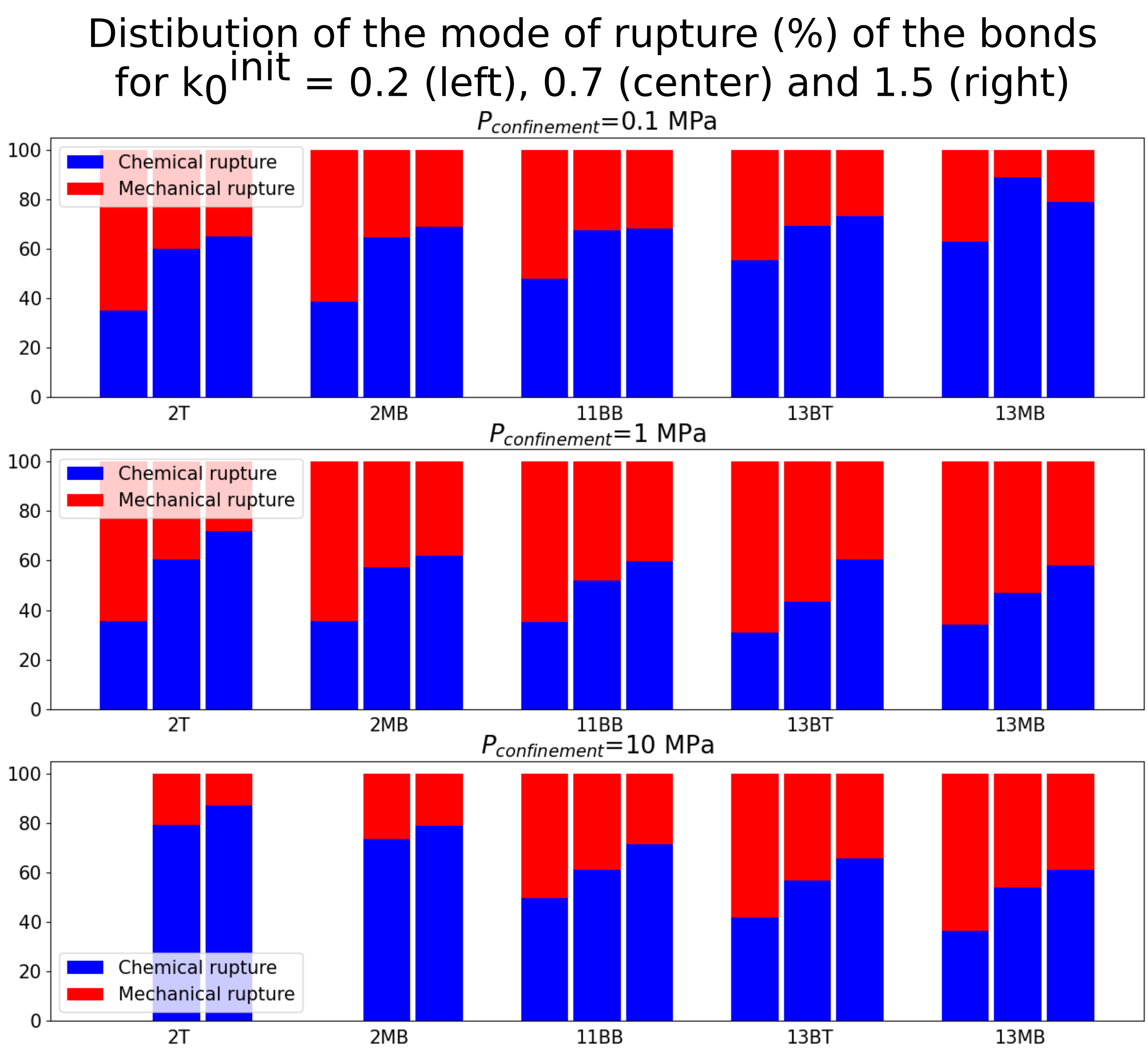}
    \caption{Distribution of the mode of rupture of the bonds at the end of the test. A bond can break by chemical dissolution (the bond surface reaches a negative or a null value) or by mechanical loading (criteria presented in Equation \ref{Bond criteria} are reached).}
    \label{Distribution Bond Rupture}
\end{figure}

The distribution of the mechanisms of bond rupture, illustrated in Figure~\ref{Distribution Bond Rupture}, highlights the diverse stress reorganization mechanisms. 
This distribution considers only the bond rupture after the sample initialization described in Section \ref{Numerical model}.
Indeed, as explained in Section \ref{Numerical model}, this preparation step induces bond breakage before the commencement of the bond dissolution, in a manner analogous to that observed in experiments.  
The bonds may break in two modes: i) by dissolution (if the bond surface $A_b$ reaches a null or a negative value) or ii) by mechanical loading (if criteria presented in Equation \ref{Bond criteria} are reached). 
It can be observed that, for a given cementation and confining pressure, the percentage of rupture by loading is greater for $k_0^{init} \leq k_0^{attr}$ than for $k_0^{init} \geq k_0^{attr}$.
To gain insight into this discrepancy, the chain forces are classified into two categories: the unstable chain force and the stable chain force, see Figure \ref{Stable Unstable CF}. 

An unstable chain force is maintained in position by the cementation at the contacts. 
Once the cement has undergone partial dissolution, the chain force collapses. It should be noted that while the particle positions remain largely unchanged, minor fluctuations induce the evolution of the chain force, resulting in the breaking of the reduced but existing cemented contacts. These unstable chain forces induce the bond rupture by mechanical loading.
In contrast, a stable chain force remains in place due to the particle-particle contact. In the event of the total dissolution of the cemented contacts, the chain force does not collapse. These stable chain forces induce the bond rupture by chemical loading.
It has been demonstrated in the previous Section that grain reorganization will only occur in the case $k_0^{init} \leq k^{attr}_0$ if the cemented contacts are reduced enough. Indeed, the unstable chain forces are predominant in this case. For an illustration of this, one may refer to the $k_0$ evolution when the Young modulus $E_m$ remains constant during the debonding process, Figure \ref{k0 evolution vs Young Modulus Assumption}a.
Even if the Young modulus remains constant during the debonding, $k_0$ evolves/granular reorganization occurs.
Once the cemented contacts have been reduced, the cementation at the contacts breaks by mechanical loading, see criterion in Equation \ref{Bond criteria}. 
Similarly, it has been demonstrated in the previous Section that grain reorganization occurs directly in the case $k_0^{init}\geq k_0^{attr}$, without the need for a notable reduction in cemented contacts.
Indeed, the stable chain forces are predominant in this case.
This is evidenced by the constant $k_0$ when the Young modulus $E_m$ remains constant during the debonding process, Figure \ref{k0 evolution vs Young Modulus Assumption}b.
As the Young modulus remains constant during the debonding, $k_0$ remains constant/granular reorganization does not occur. The cemented contacts break with a chemical rupture, as they are not loaded. The evolution of $k_0$ is a consequence of the softening of the grains. While the Young modulus decreases with the debonding process, the overlap between the grains increases, inducing grain reorganization (even if the particle positions remain largely unchanged) and a corresponding evolution of $k_0$.

Once the two mechanisms have been described, it is crucial to note that in the cases of light cementations (2T or 2MB), the final values of the $k_0$ are larger than the others if $k_0^{init}\geq k_0^{attr}$, see Figure \ref{k0 attractor Figure}. In this case, the reduction of $k_0$ is attributable to the softening of the grains. Considering light cementations, the softening is less pronounced given that the Young modulus varies between $300$ MPa (for 2T) or $320$ MPa (for 2MB) and $80$ MPa (for uncemented material). This is, for example, in contrast to $1000$ MPa (for 13MB). In the case of light cementations, the Young modulus remains approximately constant, thereby precluding the grain reorganization. The impact of the initial Young modulus value is examined in the following Section, see Figure \ref{k0 evolution vs Initial Young Modulus}. It appears that the Young modulus exerts a notable influence, particularly in the case $k_0^{init}\geq k_0^{attr}$. This influence is less pronounced in the case $k_0^{init}\leq k_0^{attr}$.

%%===========================%%

\subsection{Influence of the cementation}

Following the definition of the different degrees of cementation used by \textit{Sarkis et al.} \cite{Sarkis2022}, it appears that the percentage of contacts cemented $p_c$, the Young modulus $E_m$, and the bond sizes distribution (defined by $m_{log}$ and $s_{log}$) change with the cementation. To understand the impact of the cementation on the problem these three effects must be isolated.

\vskip\baselineskip

\begin{figure}[ht]
    \centering
    \includegraphics[width=0.6\linewidth]{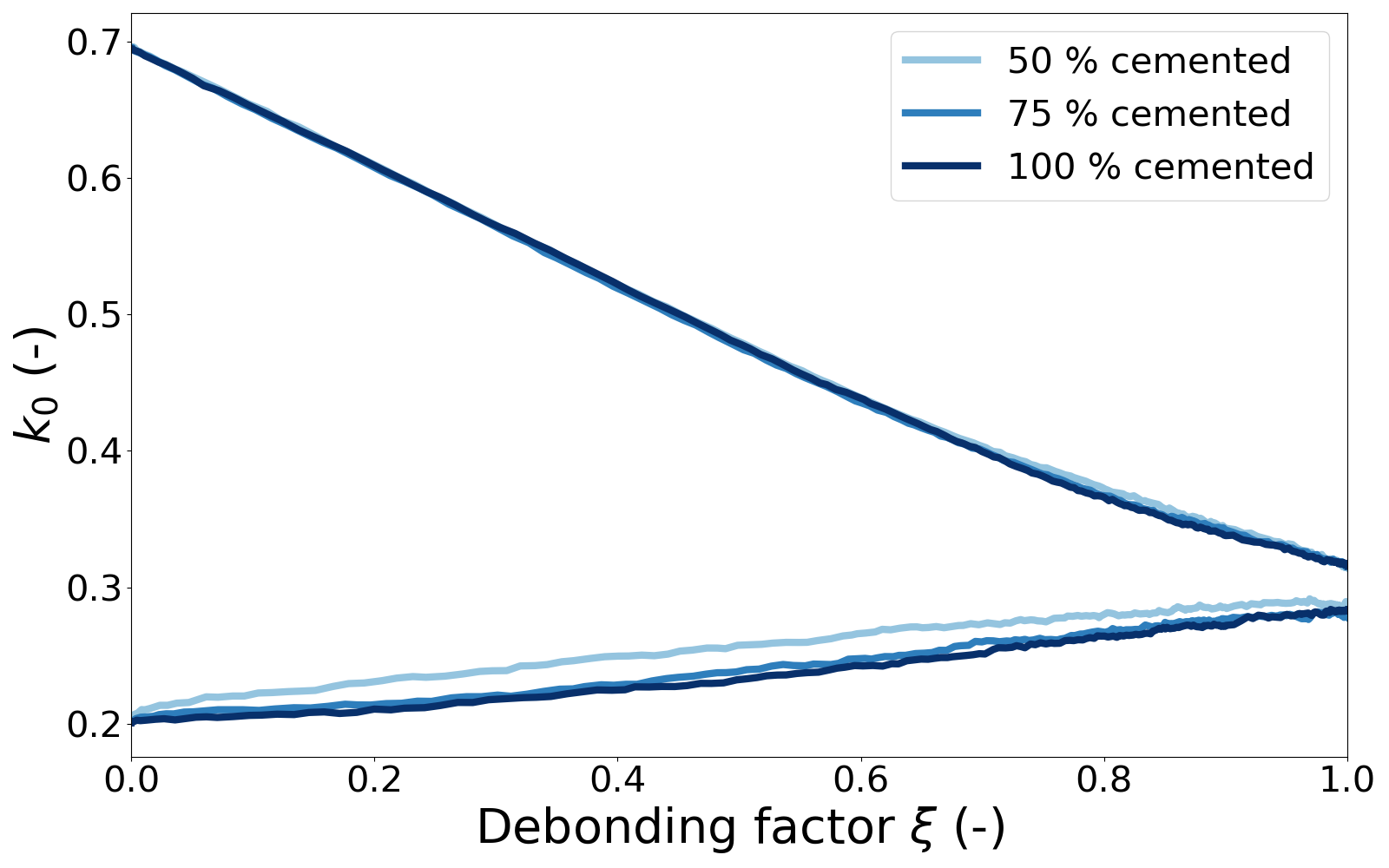}
    \caption{Evolution of the $k_0$ with the debonding variable $\xi$ for an 11BB sample under $P_{confinement}=0.1$ MPa with $50$, 75 or $100$\% initial contacts cemented.}
    \label{Result 11BB Different pc}
\end{figure}

The initial stage of the analysis focuses on the impact of the percentage of contacts that are cemented $p_c$.
This parameter describes the ratio of contact cemented during the initial configuration (see step f in Figure \ref{Initial condition algorithm}) relative to the total number of contacts. Additional simulations were conducted for comparative purposes. An 11BB sample is generated under $P_{confinement} = 0.1$ MPa with an initial $k_0=0.2$ or $0.7$. Three distinct values are considered for the parameter $p_c=50$, $75$ or $100\%$. The results are presented in Figure \ref{Result 11BB Different pc}.

It seems that $p_c$ does not affect the evolution of $k_0$. A slight discrepancy is observable for $k_0^{init}=0.2$. However, it remains negligible in comparison to the discrepancies observed with the other cementations in Figure \ref{k0 attractor Figure}. 

This discrepancy can be attributed to the differing reorganization mechanisms at stake, depending on $k_0^{init}$. As previously outlined in Section \ref{Grain Reorganization Mechanism Section}, if $k_0^{init}\leq k_0^{attr}$ the main reorganization mechanism is the collapse of the unstable chain forces. In this case, the small amount of cemented contact ($p_c$ smaller) can weaken the global structure and exert a minor influence on the $k_0$ evolution. In the case $k_0^{init}\geq k_0^{attr}$ the main reorganization mechanism is grain softening. Here, the distribution of the cemented contact has no impact. In conclusion, it can be stated that the parameter $p_c$ has no influence on the evolution of the $k_0$ during the debonding process.

\vskip\baselineskip

\begin{figure}[ht]
    \centering
    a) \includegraphics[width=0.45\textwidth]{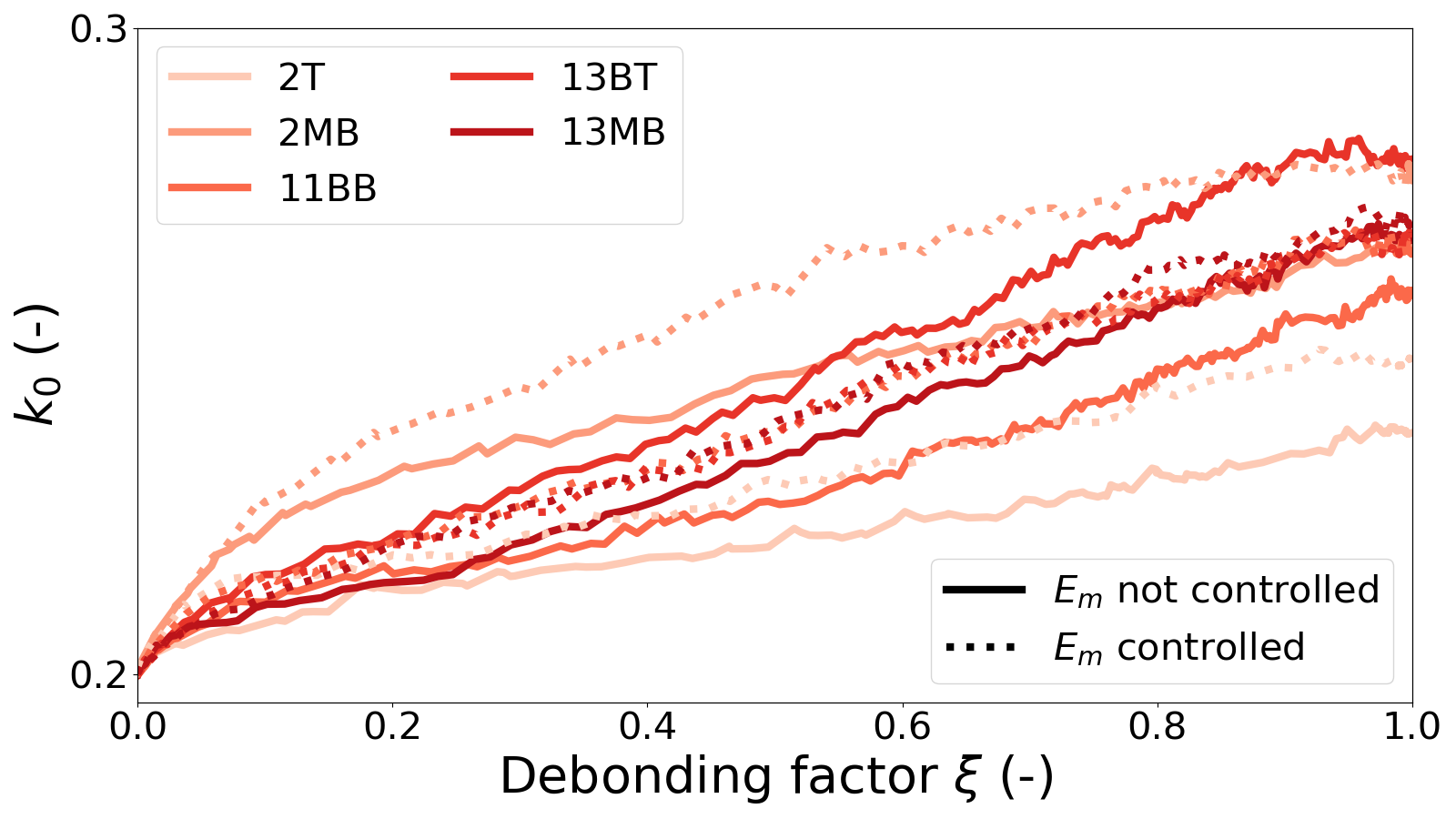}~ 
    b) \includegraphics[width=0.45\textwidth]{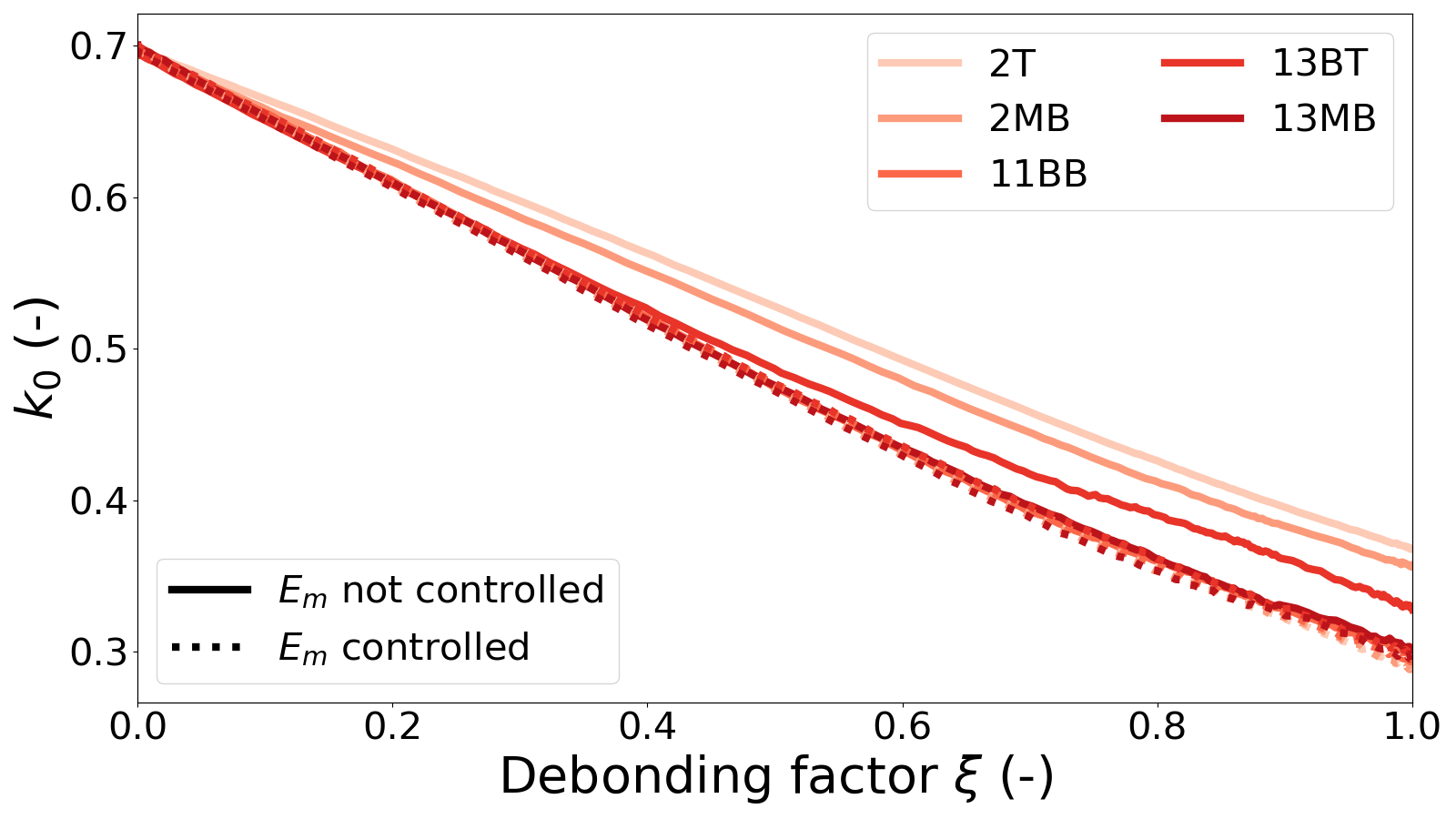}
    \caption{Evolution of the $k_0$ with the debonding factor $\xi$ for different cementations at a confinement pressure $P_{confinement}=1e6$ Pa with an initial $k_0=$~a)~$0.2$ or b)~$0.7$. The initial value of the Young modulus $E_m$ is equal to $1000$ MPa for the controlled simulations.} 
    \label{k0 evolution vs Initial Young Modulus}
\end{figure}

The second parameter to be analyzed is the Young modulus $E_m$. To investigate the effect of this parameter, simulations similar to the results presented previously are conducted, but the initial value of the Young modulus is controlled (considered the same for all cementation: $E_m = 1$ GPa, Young modulus of the cementation 13MB). The results are presented in Figure \ref{k0 evolution vs Initial Young Modulus}.

For $k_0^{init}\leq k_0^{attr}$, Figure \ref{k0 evolution vs Initial Young Modulus}a, the controlled $k_0$ evolutions appear to be approximately equivalent to the $k_0$ evolution obtained with a not controlled 13MB cementation. This can be attributed to the fact that the initial Young modulus value is identical to the one employed for this cementation. It is crucial to note that the controlled 2T and 2MB simulations give disparate results. As explained in Section \ref{Grain Reorganization Mechanism Section}, the evolution of the $k_0$ is predominantly influenced by the collapse of unstable chain forces for $k_0^{init}\leq k_0^{attr}$, rather than the softening of the grains. The presence of fewer bonds between the grains in lightly cemented samples facilitates the collapse of chain forces. Therefore, it can be concluded that the initial Young modulus has an impact on the evolution of $k_0$, although it is not the most significant factor.

For $k_0^{init}\geq k_0^{attr}$, Figure \ref{k0 evolution vs Initial Young Modulus}b, the controlled $k_0$ evolutions appear to be approximately equivalent to the $k_0$ evolution obtained with a not controlled 13MB cementation. This can be attributed to the fact that the initial Young modulus value is identical to the one employed for this cementation. Furthermore, the controlled 2T and 2MB simulations result in identical evolutions. As explained in Section \ref{Grain Reorganization Mechanism Section}, the evolution of the $k_0$ is predominantly influenced by the softening of the grains for $k_0^{init}\geq  k_0^{attr}$, rather than the unstable chain forces collapse. The grain reorganization depends significantly on the initial Young modulus value.

\vskip\baselineskip

\begin{figure}[ht]
    \centering
    \includegraphics[width=0.6\textwidth]{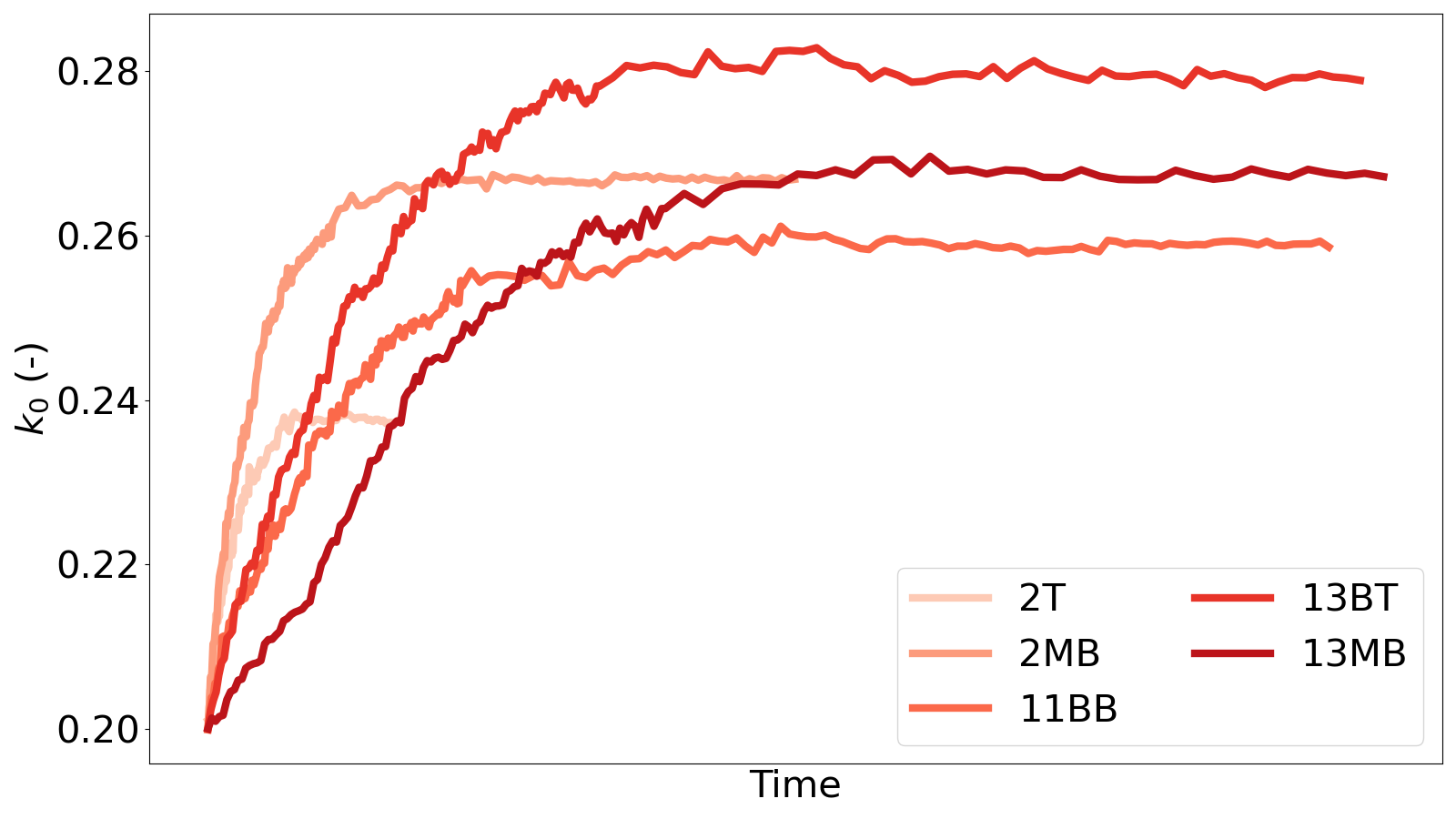}
    \caption{Evolution of the $k_0$ with the cumulative bond surface dissolved ($\approx$ the time) for the different cementations at $P_{confinement}=1$ MPa.}
    \label{Influence BSD Figure}
\end{figure}

The third parameter to be analyzed is the bond size distribution, which is controlled by $m_{log}$ and $s_{log}$, see Appendix \ref{Lognormal Distribution}. As previously discussed in this Section, the bond size distribution affects the case $k_0^{init}\leq k_0^{attr}$, Figure \ref{k0 evolution vs Initial Young Modulus}a, but not in the case $k_0^{init}\geq k_0^{attr}$, Figure \ref{k0 evolution vs Initial Young Modulus}b. 
As previously presented in Section \ref{Grain Reorganization Mechanism Section}, the mechanism of the $k_0$ evolution can be attributed to the collapse of the unstable chain forces if $k_0^{init} \leq k_0^{attr}$. It is important to note that while the positions of the particles remain largely constant, minor fluctuations can induce changes in the chain forces. The bond size distribution exerts a notable influence on this phenomenon, particularly the variance of the distribution $s_{log}$.
Furthermore, Figure \ref{Influence BSD Figure} illustrates that the light cementations (2T and 2MB) undergo complete dissolution at a significantly faster rate than the high cementations (13BT and 13MB). This Figure represents the evolution of the $k_0$ with the cumulative bond surface dissolved (similar to the times). The mean of the distribution $m_{log}$ appears to control the temporal aspect of the debonding phenomena.

%%===========================%%

\subsection{Influence of the confinement}

\begin{figure}[ht!]
    \centering
    a) \includegraphics[width=0.8\textwidth]{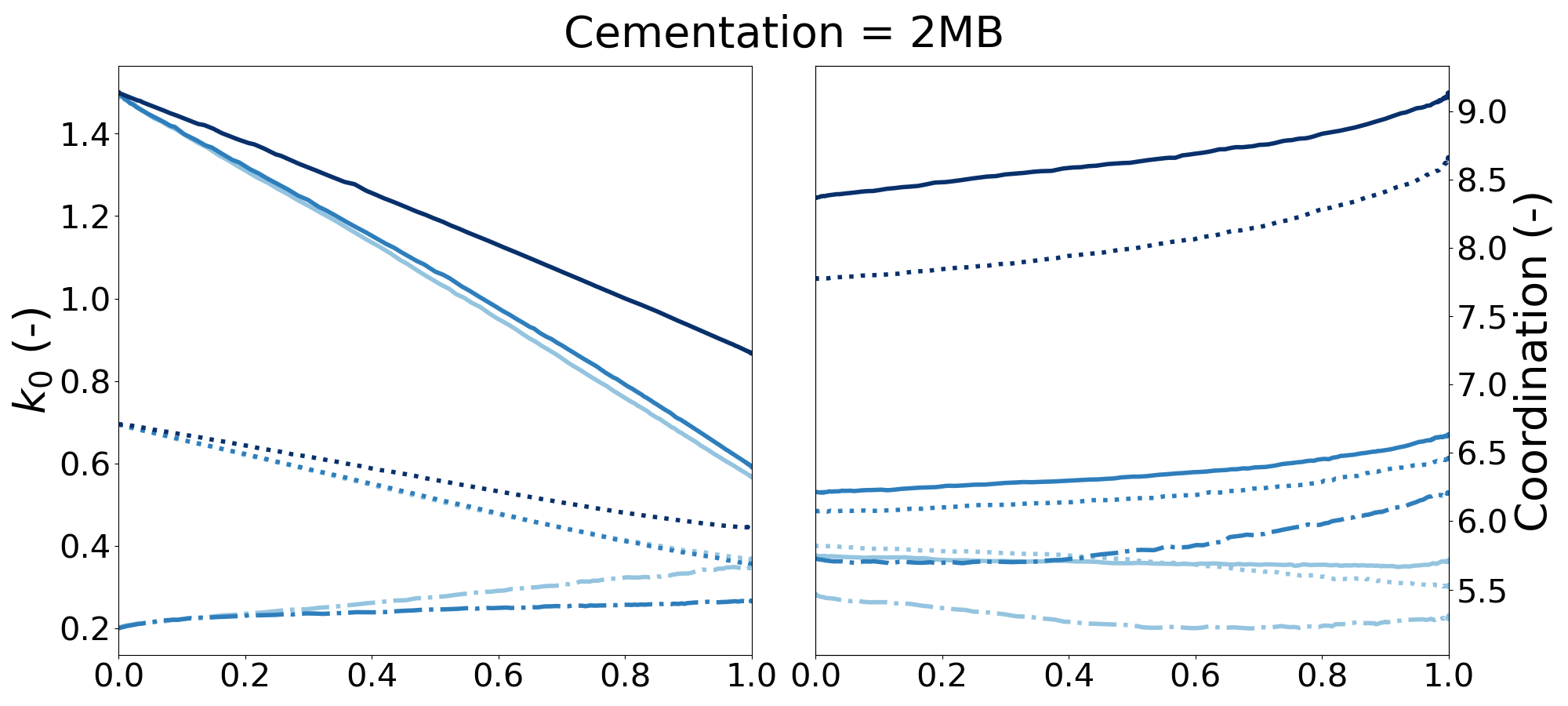}\\
    b) \includegraphics[width=0.8\textwidth]{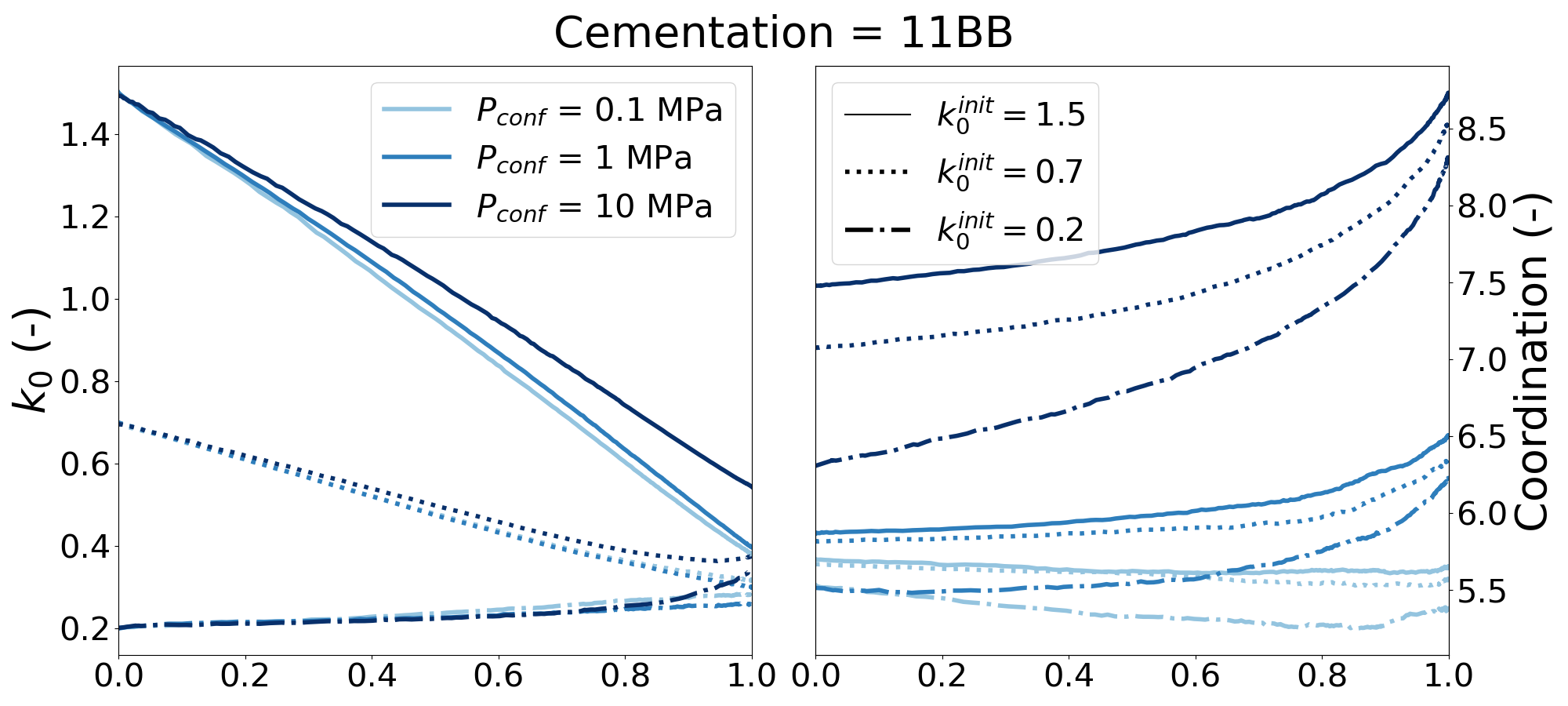}\\
    c) \includegraphics[width=0.8\textwidth]{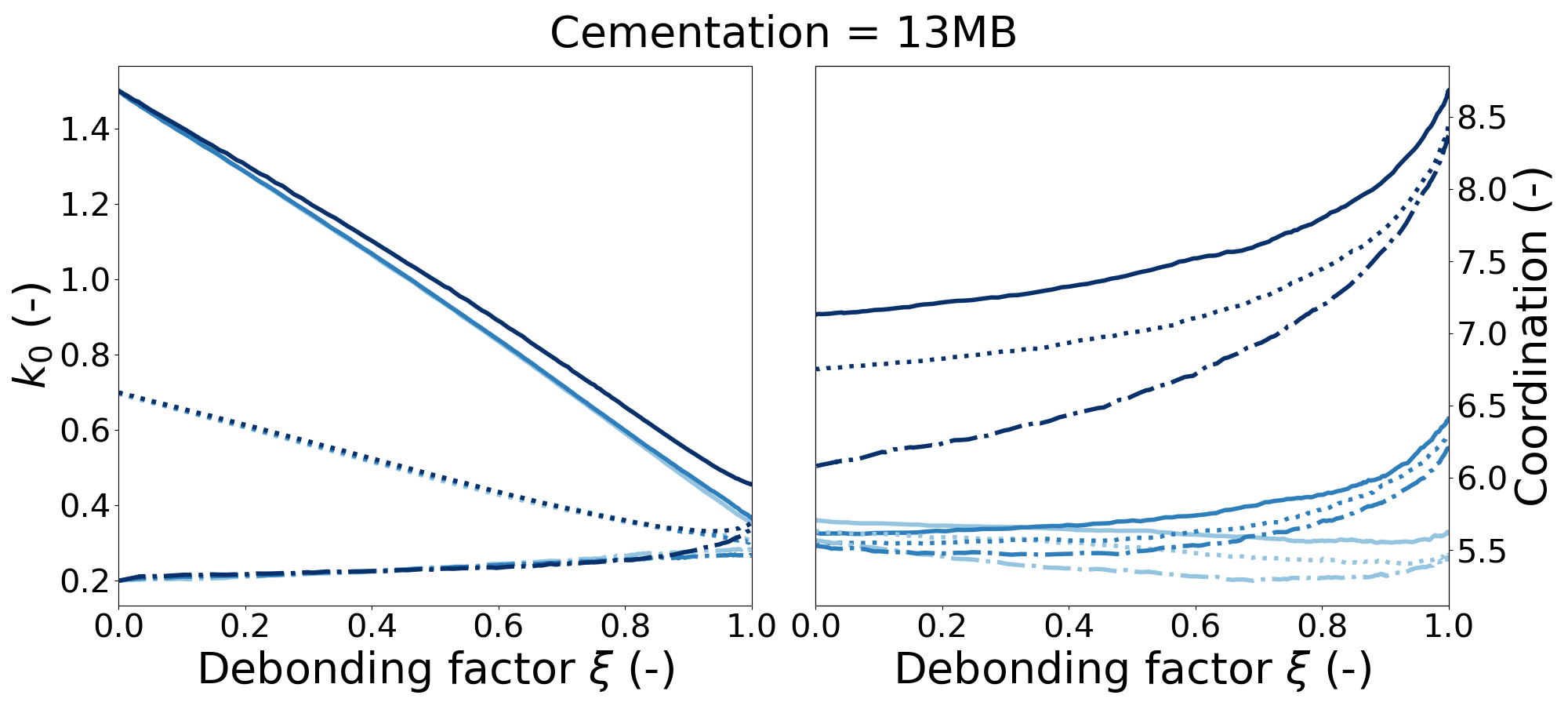}
    \caption{Evolution of the $k_0$ and the coordination number with the debonding variable $\xi$ at $P_{confinement}=0.1$, $1$ and $10$ MPa, with $k_0^{init}=0.2$, $0.7$ and $1.5$ for three different degrees of cementation: a) 2MB, b) 11BB and c) 13MB.}
    \label{Influence P_confinement}
\end{figure}

The impact of the confinement pressure $P_{confinement}$ is studied in this Section. The results are shown in Figure \ref{Influence P_confinement}.

It should first be noted that no results can be obtained for the light cementations (2T and 2MB) at $P_{confinement}=10$ MPa when $k_0^{init}=0.2$. The bonds are found to break during the initial configuration set-up, due to the high confinement pressure. As previously explained in Section \ref{Numerical model}, $k_0$ is around $k_0^{attr}$ at the step g of Figure \ref{Initial condition algorithm}. In order to obtain $k_0^{init}\leq k_0^{attr}$, it is necessary to extend the lateral dimension of the sample. The combination of this extension with the application of the vertical pressure at the top results in bond breakage. This phenomenon occurs less frequently in the case $k_0^{init}\geq k_0^{attr}$ (compression of the lateral dimension of the sample) because the bond breakage occurs only under shear or a tensile conditions, see Equation \ref{Bond criteria}. Indeed, the sample is not under tensile loading and the shear condition occurs more frequently in extension ($k_0^{init}\leq k_0^{attr}$) than in compression ($k_0^{init}\geq k_0^{attr}$).

Figure \ref{Influence P_confinement} illustrates that the coordination number (the mean number of contacts per grain) increases with the confining pressure $P_{confinement}$. As the confining pressure increases, the grains are more squeezed and the number of contacts increases. In DEM, this phenomenon requires careful consideration, particularly in the case of rigid particles. Indeed, the models in question are built on the basis of the small overlap assumption. This limit is named the jammed state \cite{Liu1998}. As illustrated in the Figure, the grain reorganization can be divided into two distinct phases: first a linear part, then a nonlinear. It should be noted that the Young modulus of the grains $E_m$ reduces as a result of debonding, which serves to indicate that the sample is progressing towards the jammed state. Given that the Young modulus is smaller for lighter cementations, this state is reached sooner.
The examples $P_{confinement}=10$ MPa (green lines) and $k_0^{init}=1.5$ (plain lines) illustrate this aspect. The nonlinear portion of the curve begins to emerge at approximately $\xi=0.6$ for an 11BB sample and $\xi=0.7$ for a 13MB sample. In a similar idea, the final value of the $k_0$ is globally larger when $P_{confinement}$ is larger for $k_0^{init} \geq k_0^{attr}$. As the sample approaches the jammed state, the available space for reorganization is reduced. This observation is less evident in the case $k_0^{init} \leq k_0^{attr}$. Indeed, the mechanism of the $k_0$ evolution is dominated by the collapse of the unstable chain forces, rather than the softening of the grain, see Section \ref{Grain Reorganization Mechanism Section}.

%%=======================================================%%

\section{Discussion}

%%===========================%%

\subsection{The evolution of stress state can generate failure}

\begin{figure}[ht]
    \centering
    a) \includegraphics[width=0.45\textwidth]{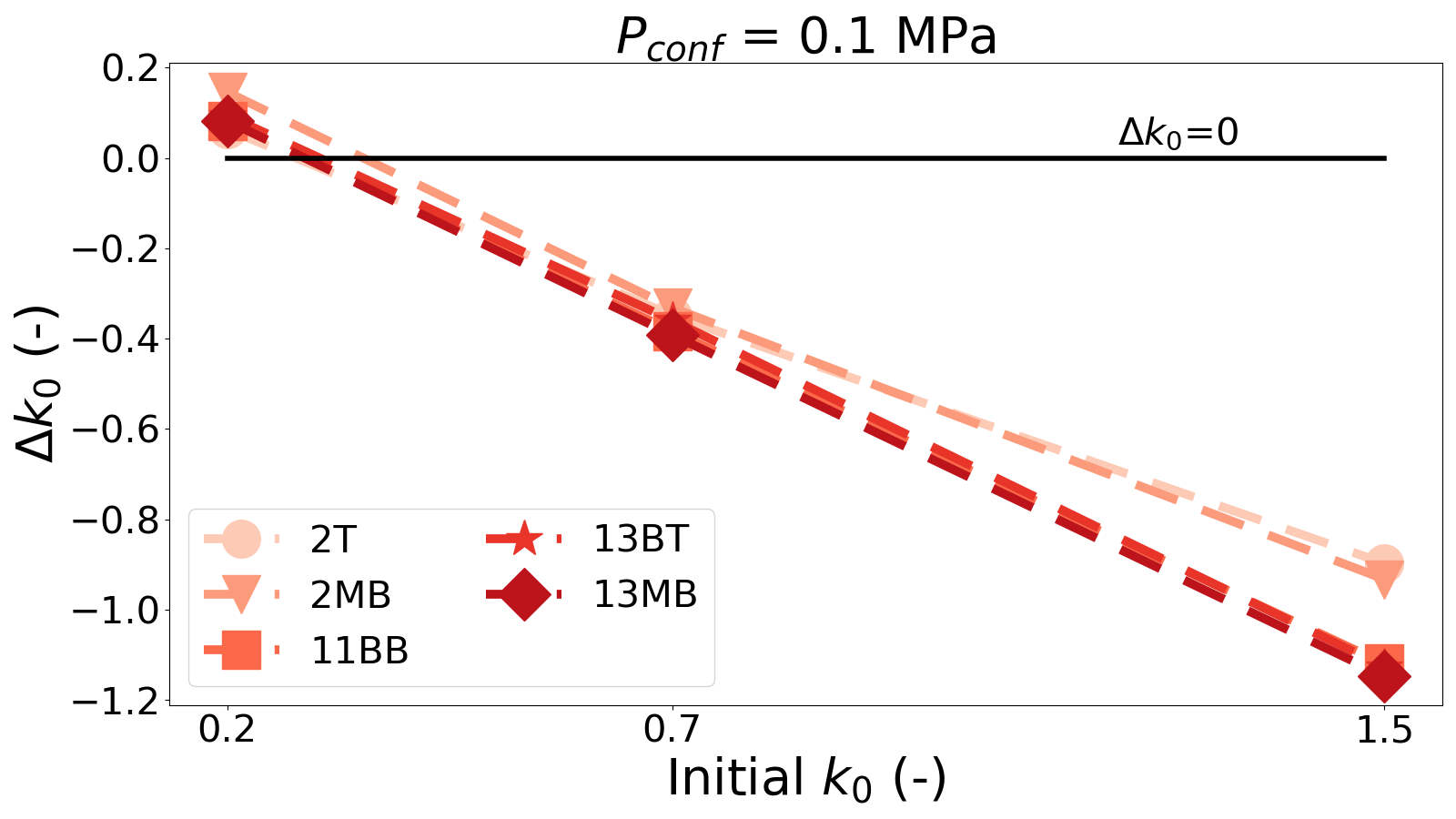}
    b) \includegraphics[width=0.45\textwidth]{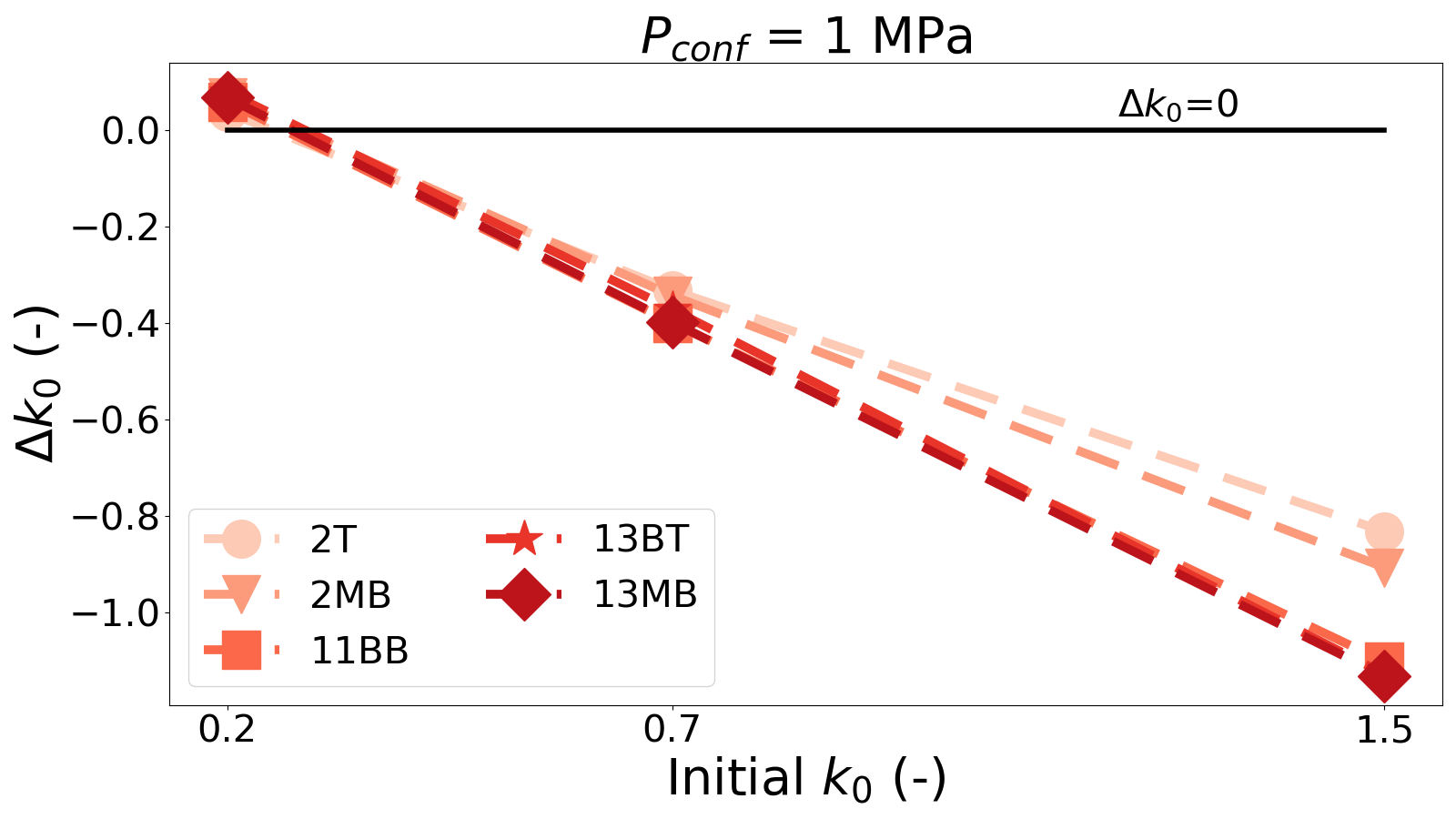}\\
    c) \includegraphics[width=0.45\textwidth]{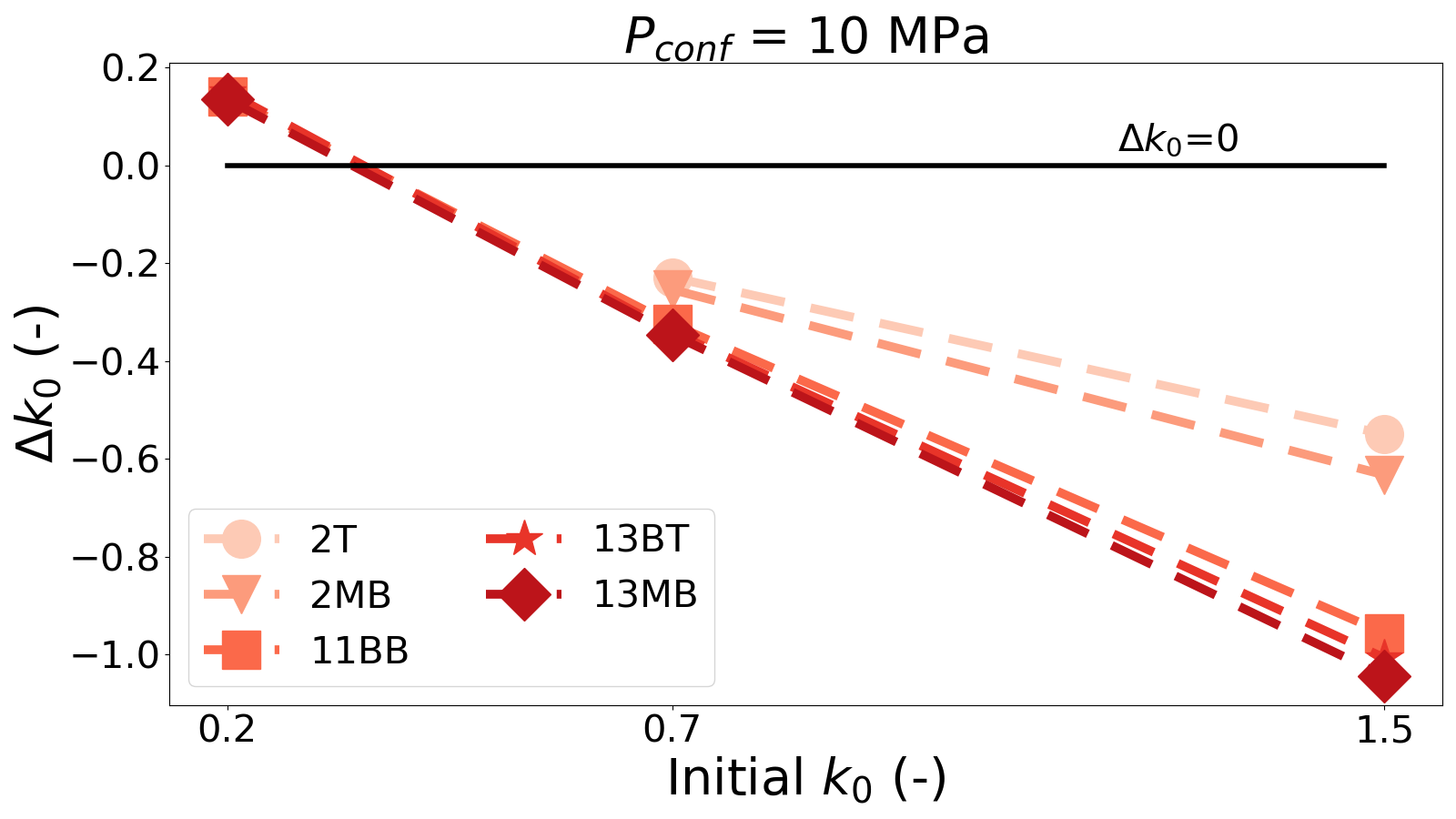}
    \caption{Evolution of the $k_0$ between the initial and final configurations $\left(\Delta k_0=k_0^{final}-k_0^{initial}\right)$ for different initial $k_0^{initial}$ values and all the cementations at different confinement pressures $P_{confinement}=$~a)~$0.1$~MPa b)~$1$~MPa or c)~$10$~MPa.}
    \label{dk0 plots}
\end{figure}

Figure \ref{dk0 plots} shows the evolution of $k_0$ $\left(\Delta k_0=k_0^{final}-k_0^{initial}\right)$ between the initial and final configurations for different initial $k_0^{initial}$ values at different confinement pressures $P_{confinement}$ and for all the degrees of cementation. 

\newpage

\begin{figure}[ht]
    \centering
    \includegraphics[width=0.6\textwidth]{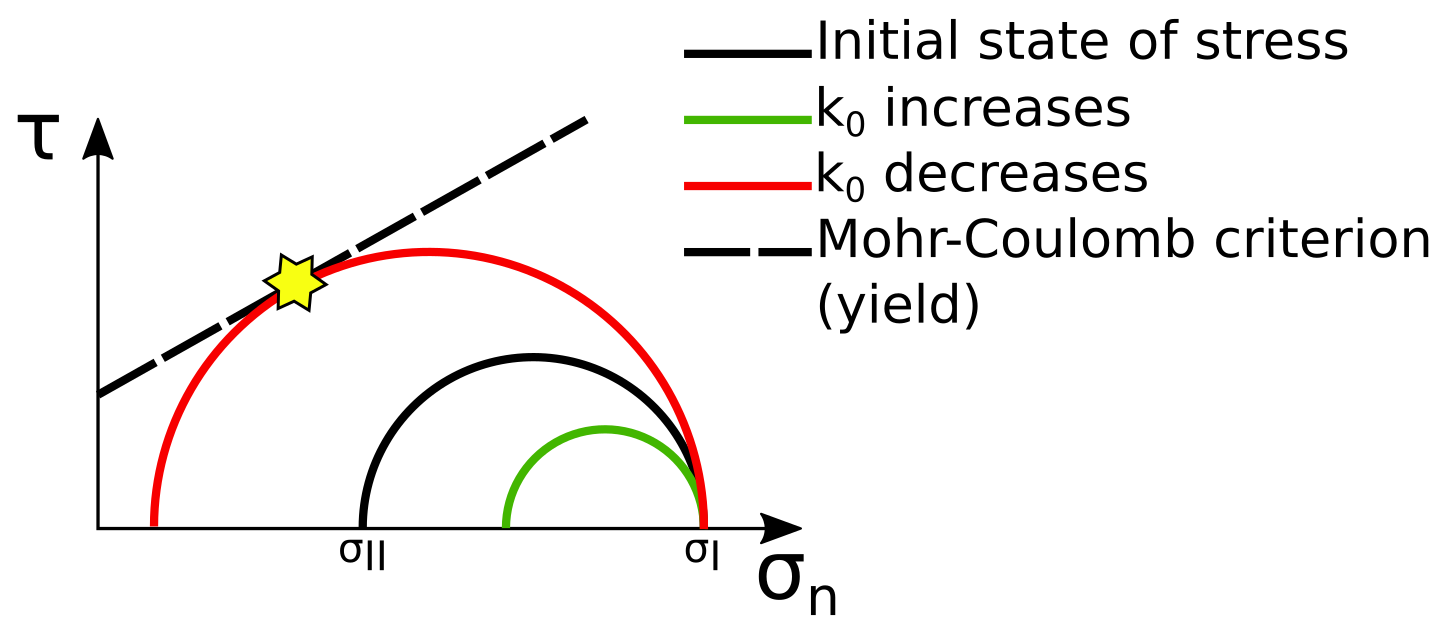}
    \caption{The effect of an increase or a decrease of the $k_0$ in the Mohr plane, a reduction of this parameter can induce a failure.}
    \label{Mohr Coulomb Fault}
\end{figure}

The evolution of the $k_0$ during dissolution can be of significant consequence, particularly in the context of an underground reservoir. The host rock or the triggering of slip along an existing fault can be modeled using a simple Mohr-Coulomb criterion, illustrated in Figure \ref{Mohr Coulomb Fault}. It appears that a reduction of $k_0 \left(=\sigma_{II}/\sigma_I\right)$ can result in a failure. Indeed, the diameter of the Mohr circle increases $\left(D_{MC} = \sigma_I\left(1-k_0\right)\right)$, where $D_{MC}$ is the diameter of the Mohr circle and $\sigma_I$ is the vertical stress and remains constant \cite{Shin2008}. As previously demonstrated in this study, the $k_0$ reduction appears only in the case $k_0^{init} \geq k_0^{attr}$. Therefore, this evolution of the $k_0$ can potentially induce the nucleation of slip, which can lead to an earthquake if the fault is seismogenic \cite{Keranen2018}.

%%===========================%%

\subsection{On the Young modulus reduction assumption}
\label{Discussion Young modulus reduction}

As explained in Section \ref{Young modulus Reduction Results} and illustrated in Figure \ref{k0 evolution vs Young Modulus Assumption}, the assumption of a Young modulus reduction with the debonding (Equation \ref{Young Reduction}) is pivotal for the evolution of the stress state of a sample subjected to dissolution of the bonds, particularly in the case of $k_0^{init}\geq k_0^{attr}$. It has been demonstrated that $k_0$ evolves only when the Young modulus is reduced with the debonding process. This assumption on the evolution of the Young modulus represents not only a numerical investigation of the phenomenon of dissolution but also the different types of contacts and loading history of the sample.

As illustrated in Figure \ref{Type contact} and detailed in Section \ref{Numerical model}, a cemented contact can be categorized into three distinct types: I) Frictional, II) Mixed/Cohesive (Frictional+Cemented), or III) Cemented \cite{Sarkis2022}. For type I, there is no cement, so the Young modulus reduction assumption is not required. Otherwise, the contact can be modeled as two springs (one for the grains and one for the cement) in parallel (type II) or in series (type III). For these two types (parallel or series), the Young modulus reduction assumption should be employed, contingent on the loading history, as discussed in the following paragraph. The distinction between type II and III hinges on the ultimate value of the Young modulus $E_m^{final}$, post-complete dissolution of the bonds. For type II, the ultimate value will be the one which is assumed for frictional contact (type I) $E_m^{I}$. For type III, this value will be zero (unless a new contact of type I occurs).

As explained earlier in this Section, the Young modulus reduction assumption can be applied to Type II and III contacts, depending on the loading history.
For the purposes of this discussion, it is assumed that cementation is initially generated at pressure $P_i$. The sample can be modeled as springs (one for the grains, one for the bonds) in parallel (Type II) or series (Type III). The cementation is considered loaded only when the applied pressure $P_j$ verifies $P_j>P_i$. 
Subsequently, the stiffness of the bond spring reduces as the dissolution process occurs. The Young modulus reduction assumption is applied to ensure the sample is representative. 
An illustration of this configuration is provided in Figure \ref{k0 attractor Figure}. It seems that $k_0$ evolves to reach the value $k_0^{attr}$.
If $P_j=P_i$, the cementation is unloaded, and the debonding has no impact on the mechanical behavior of the sample. It would be erroneous to apply the Young modulus reduction assumption. An illustration of this configuration is provided in Figure \ref{k0 evolution vs Young Modulus Assumption}. It seems that $k_0$ reaches the value $k_0^{attr}$ only in the case $k_0^{init} \leq k_0^{attr}$ and remains constant in the case $k_0^{init} \geq k_0^{attr}$. This preceding discussion is summarized in Table \ref{Summarize Young Reduction}.

\begin{table}[ht]
    \centering
    \begin{tabular}{|l|c|c|c|}
         \multicolumn{1}{c|}{}&Contact I&Contact II&Contact III \\
         \hline
         Bonds loaded&NO&YES ($E_m^{final}=E_m^{I})$&YES ($E_m^{final}=0)$ \\
         \hline
         Bonds not loaded&NO&NO&NO \\
         \hline
    \end{tabular}
    \caption{Summarize the use of the Young modulus reduction assumption. If the hypothesis is applied, the final value of $E_m$ is given.}
    \label{Summarize Young Reduction}
\end{table}

%%===========================%%

\subsection{A Young modulus dependency on the size of the bonds}
\label{Young from Ab}

As previously mentioned in Section \ref{Numerical model}, the relationship described in Equation \ref{Young Reduction} between the contact scale ($E_m$) and the sample scale ($\xi$) could be a matter of contention. 
This Section explores an alternative approach, in which the local stiffness at each contact is determined by the bond surface $E_m(A_b)$ \cite{Sun2016,Sun2018}, see Equation \ref{Young Reduction Local}. This alternative approach eliminates the discussed relationship between the contact ($E_m$) and sample ($\xi$) scales.

\begin{equation}
    E_m(A_b) = \frac{E_m^{cementation}-E_m^{untreated}}{A_{b,\,ref}^{cementation}}\times A_b + E_m^{untreated}
    \label{Young Reduction Local}
\end{equation}
where $E_m$ is the Young modulus of the contact, $E_m^{cementation}$ is the initial Young modulus depending on the initial degree of cementation, see Table \ref{Parameters Used Cementation}, $E_m^{untreated}$ is the Young modulus without cementation (here $80$ MPa), $A_{b,\,ref}^{cementation}$ is a bond surface of reference depending on the initial degree of cementation, and $A_b$ is the bond surface of the contact.
Each cementation presented in Table \ref{Parameters Used} is represented through the parameters $E_m^{cementation}$ and $A_{b,\,ref}^{cementation}$. This bond surface of reference  $A_{b,\,ref}^{cementation}$ is computed to ensure that the mean Young modulus within the sample after the cementation, see step f) in Figure \ref{Initial condition algorithm}, is equal to $E_m^{cementation}$ from Table \ref{Parameters Used Cementation}. 
Compared to the previous model, a distribution of the Young modulus appears within the sample. 
From the bond size distribution depicted in Appendix \ref{Lognormal Distribution} and the parameters $p_c$, $m_{log}$, and $s_{log}$ from Table \ref{Parameters Used Cementation}, the different values of $A_{b,\,ref}^{cementation}$ are available in Table \ref{Ab ref}.

\begin{table}[h]
    \centering
    \begin{tabular}{|l|c|c|c|c|c|}
    \hline
    Cementation & 2T & 2MB & 11BB & 13BT & 13MB\\
    \hline
    $A_{b,\,ref}^{cementation}\; (\mu m^2)$& 148 & 2303 & 4267 & 6577 & 8131\\
    \hline
    \end{tabular}
    \caption{Bond surface of reference with the different degrees of cementation.}
    \label{Ab ref}
\end{table}

\begin{figure}[ht]
    \centering
    a) \includegraphics[width=0.45\textwidth]{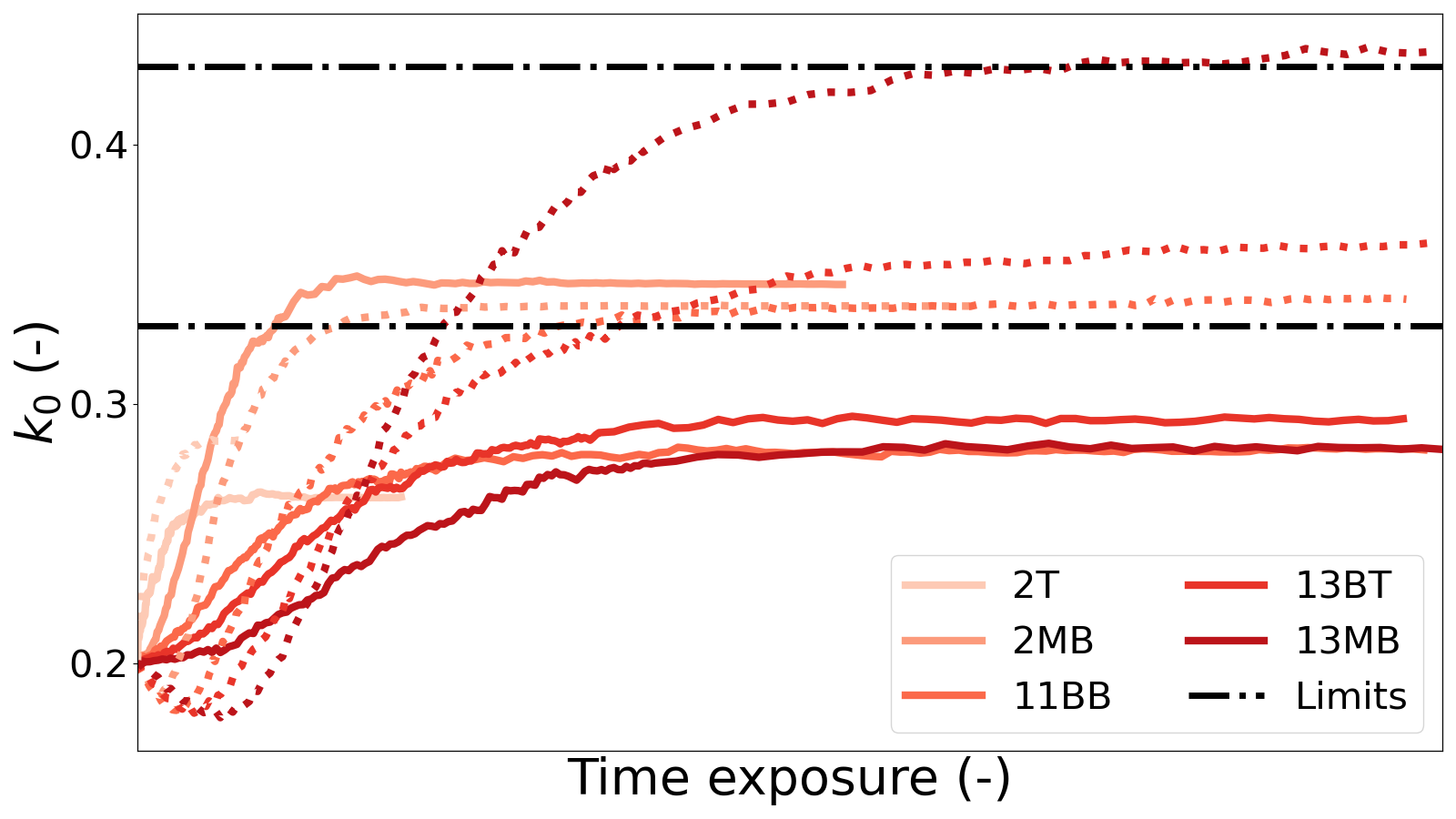} 
    b) \includegraphics[width=0.45\textwidth]{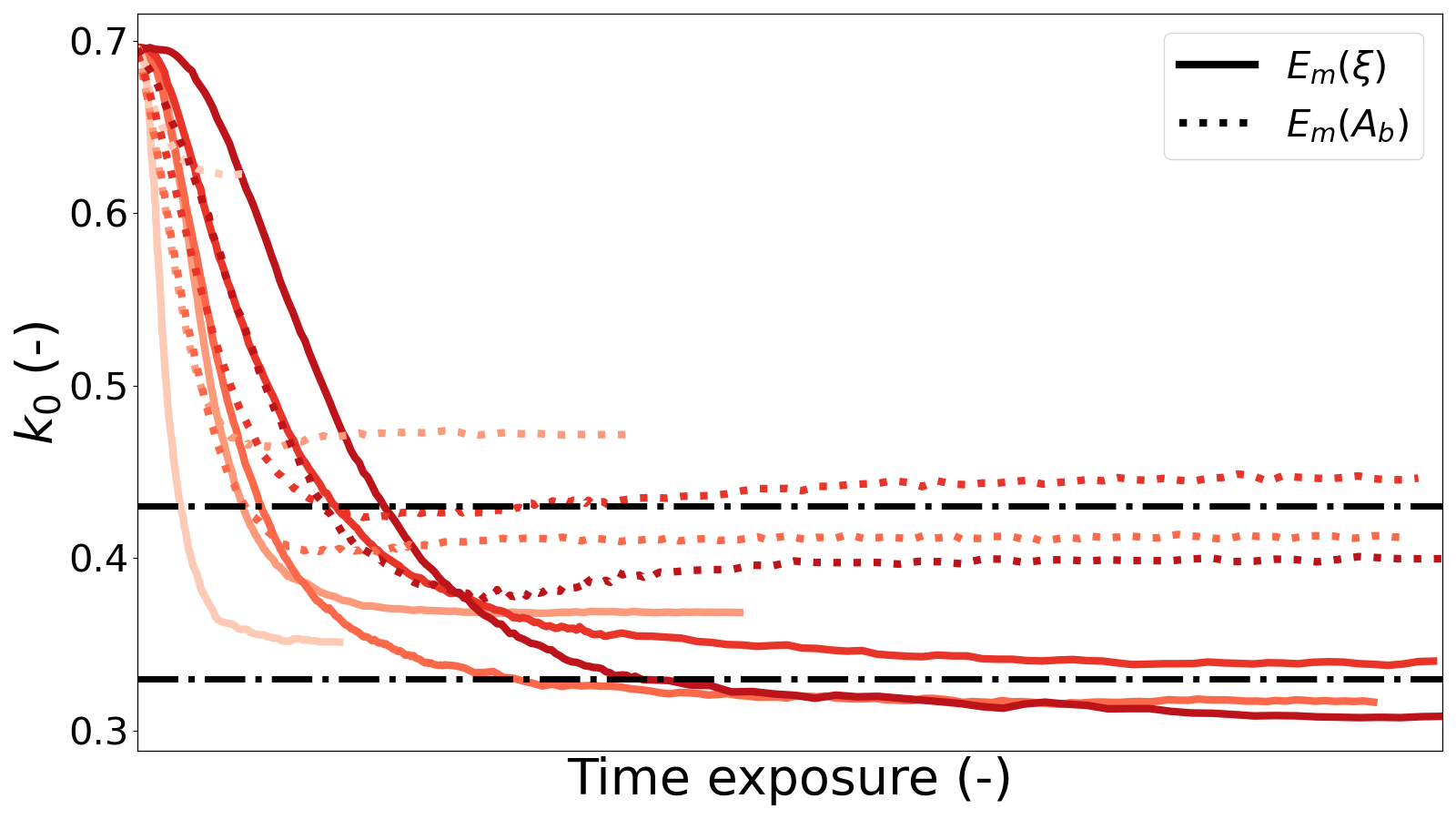} 
    \caption{Evolution of $k_0$ with the time exposure for $P_{confinement}=0.1$ MPa for $E_m(\xi)$ and $E_m(A_b)$ in the case $k_0^{init} =$ a) $0.2$ and b) $0.7$. Limits considering $k^{attr}_0 = \nu/(1-\nu)$ are plotted with $\nu = 0.25 - 0.3$.}
    \label{k0 evolution Em local vs global}
\end{figure}

Similar to Equation \ref{Bond criteria}, the contact strength still depends on the local bond surface area $A_b$. If one of the rupture criteria is reached, the Young modulus of the contact is considered equal to the Young modulus without cementation ($80$ MPa).
The results of the $E_m(\xi)$ and $E_m(A_b)$ simulations are depicted in Figure \ref{k0 evolution Em local vs global}.
Similar to Figure \ref{k0 attractor Figure}, it has been shown analytically by \cite{Vaughan1984} that $k_0$ approaches a limit given by $k_0^{attr}=\nu/(1-\nu)$ during weathering and stiffness reduction. Considering a window of $[0.25-0.3]$ for the value of $\nu$ \cite{Cheng2017,Ziccarelli2024,Mortazavi2024}, the attractor value window is $[0.33-0.43]$, in agreement with the results obtained.

\begin{figure}[h]
    \centering
    \includegraphics[width=0.6\linewidth]{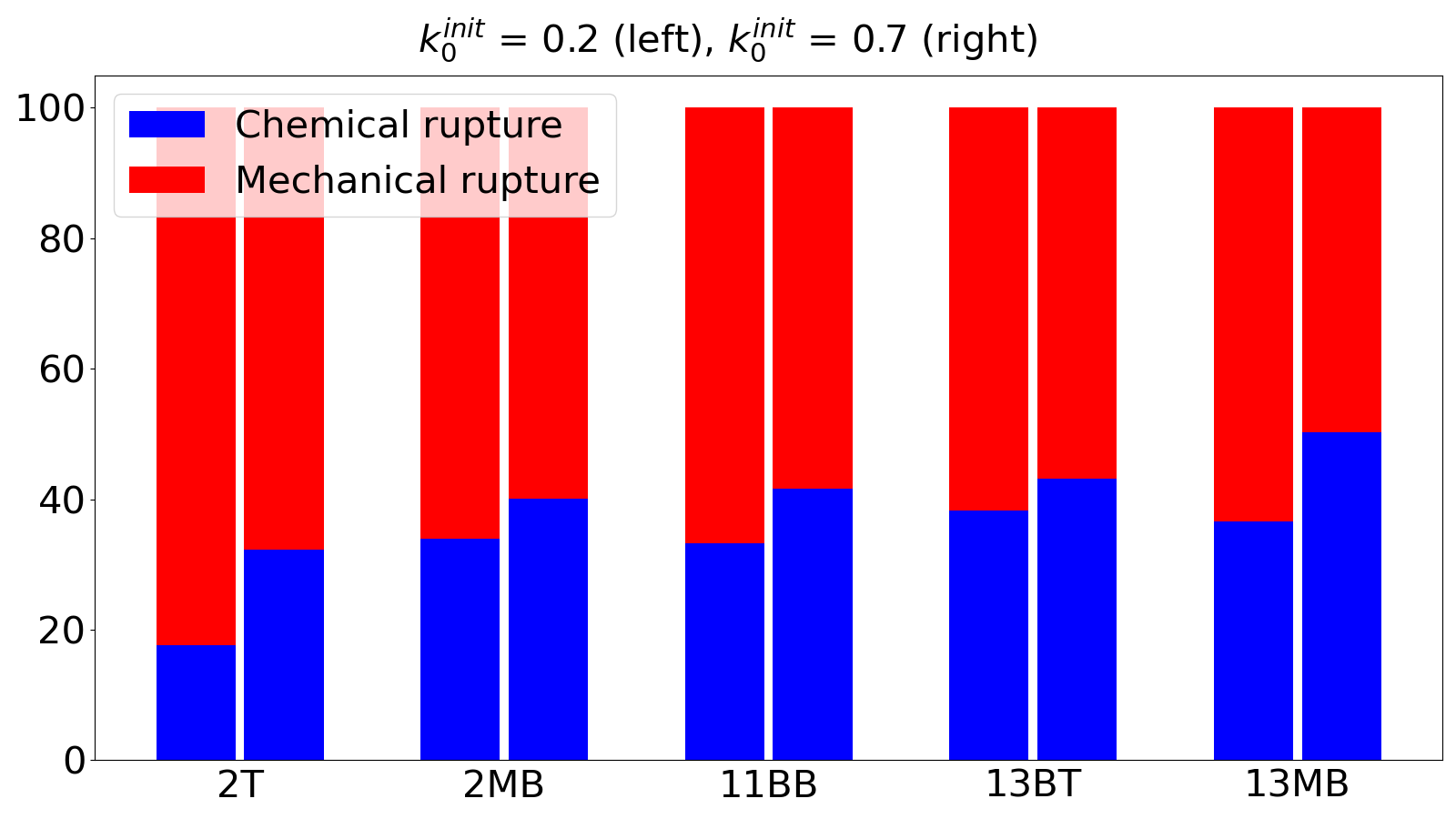}
    \caption{Distribution of the mode of rupture of the bond at the end of the test. A bond can break by chemical dissolution (the bond surface reaches a negative or a null value) or by mechanical loading (criteria presented in Equation \ref{Bond criteria} are reached).}
    \label{Rupture Mode Local}
\end{figure}

Moreover, it appears that the $k_0$ evolution mechanisms revealed for $E_m(\xi)$ in Section \ref{Grain Reorganization Mechanism Section} remain the same in the case of $E_m(A_b)$. Indeed, Figure \ref{Rupture Mode Local} emphasizes the mode of rupture of the bond at the end of the test for the distinct cementation and initial state of stress, and similarly to Figure \ref{Distribution Bond Rupture}, the bonds break more by mechanical loading in the case $k_0^{init}\leq k_0^{attr}$ than in case $k_0^{init}\geq k_0^{attr}$. 
As explained in Section \ref{Grain Reorganization Mechanism Section}, this behavior is due to the predominance of the unstable chain force in the case $k_0^{init}\leq k_0^{attr}$ and the stable chain force in the case $k_0^{init}\geq k_0^{attr}$.

Even if the attractor configuration exists and the evolution mechanisms are similar, the final states of the stress are different in the cases $E_m(\xi)$ and $E_m(A_b)$.
In the case $k_0^{init}\leq k_0^{attr}$, the final value of $k_0$ appears smaller for $E_m(\xi)$ than for $E_m(A_b)$. To understand this behavior, it is worth remembering that the particles after the bond breakage are stiffer in the case $E_m(\xi)$ than in the case $E_m(A_b)$. While $E_m$ is considered equal to $80$ MPa after the breakage of the bond in the case $E_m(A_b)$, the value considered in the case $E_m(\xi)$ is larger, see Equation \ref{Young Reduction}.
The rupture of the unstable chain forces being the $k_0$ evolution mechanism in the case  $k_0^{init}\leq k_0^{attr}$, stiffer $E_m(\xi)$ particles located on the sides limit the collapse of the unstable chain forces compared to softer $E_m(A_b)$ particles \cite{Radjai1998}, reducing the amplitude of the $k_0$ evolution.
In the same vein, the final value of $k_0$ appears larger for $E_m(A_b)$ than for $E_m(\xi)$ in the case $k_0^{init}\geq k_0^{attr}$. 
The softening of the particles being the $k_0$ evolution mechanism in the case $k_0^{init}\geq k_0^{attr}$, the $k_0$ evolution is limited in the $E_m(A_b)$ sample as $E_m$ is considered equal to $80$ MPa after the breakage of the bond, while the value considered in the $E_m(\xi)$ sample is larger, and remains decreasing, see Equation \ref{Young Reduction}. The amplitude of the $k_0$ evolution is then smaller in the case $E_m(A_b)$ than in the case $E_m(\xi)$.

The remaining question is to determine the accurate model between the assumptions $E_m(\xi)$ and $E_m(A_b)$. 
The trivial answer is to consider the case $E_m(A_b)$ as the most accurate option. Indeed, the $E_m(\xi)$ model implies that the contact scale ($E_m$) depends on the sample scale ($\xi$), which is controversial.
However, coming back to the assumption of the predominance of the type II for the contacts, see Figure \ref{Type contact}, it appears that the bond material participates in the stress transmission (not for traction), even if the contact is broken because of a mechanical loading. The Young modulus of this broken contact with residual bond material should be larger than the Young modulus considered for the uncemented grain-grain contact (type I) as the cement continues to act in parallel to the grain-grain contact and thus provide additional rigidity.
The more progressive decrease of the contact modulus with $E_m(\xi)$ allows to take into account this effect to some extent, which is not verified by the model $E_m(A_b)$. 
As often with models, the experimental reality is located between boundary scenarios (in this work, $E_m(\xi)$ and $E_m(A_b)$). 
The existence of the attractor configuration, the mechanisms inducing $k_0$ evolution, and the main consequence (fault reactivation) are similar in the boundary scenarios, only the pattern of $k_0$ evolution is impacted by the model assumptions.

%%=======================================================%%

\section{Conclusion}

The debonding of a rock during weathering can result in the material failure as a consequence of the deterioration of its mechanical properties, and the subsequent evolution of the state of stress. This effect has the potential to have significant implications for fault nucleation and reactivation in the context of underground storage projects. This paper explores the impact of several parameters such as the degree of cementation, the confining pressure, the initial value of $k_0$, and the loading history.

It has been demonstrated that an attractor configuration exists and that grain reorganization occurs during the debonding in order to reach this state. In particular, $k_0$ evolves to its attractor value, exhibiting either an increase or a decrease. It appears that a $k_0$ reduction, which is the most likely in a reservoir configuration, can be at the origin of fault reactivation and, consequently, induced seismicity.

Two main mechanisms have been identified as responsible for grain reorganization, contingent on the initial state of stress.
It is necessary to distinguish between two categories of chain forces: the unstable chain force and the stable chain force. 
The unstable chain forces provide support for the force thanks to the cementation. However, these structures collapse when the debonding occurs. In contrast, the stable chain forces remain in place even in the absence of cementation. It appears that the unstable chain forces are dominant in the case $k_0^{init}\leq k_0^{attr}$ and the stable chain forces are dominant in the case $k_0^{init}\geq k_0^{attr}$. In the latter case, grain reorganization occurs with grain softening.

%%=======================================================%%

\section*{Acknowledgements}

This research has been partially funded by the Fonds Spécial de Recherche (FSR), Wallonia-Bruxelles Federation, Belgium. The
work has also received funding from the National Science Foundation (NSF), USA, project CMMI-2042325.

%%=======================================================%%

\section*{Author Contributions}

All authors contributed to the study Conceptualization, Methodology, Writing (original draft), and Writing (review \& editing). 
Software, Data curation, Investigation, Visualization: Alexandre Sac-Morane.
Funding acquisition, Supervision: Hadrien Rattez, Manolis Veveakis.
All authors read and approved the final manuscript.

%%=======================================================%%

\appendix

\section{The lognormal distribution}
\label{Lognormal Distribution}

The probability of a bond to have a surface $A_b$ follows a lognormal distribution formulated in Equation \ref{Lognormal Distribution Equation}. This distribution is defined by the expected value $m_{log}$ and the variance $s_{log}$.

\begin{equation}
    p(A_b) = \frac{1}{A_b\,s_{log}\sqrt{2\pi}}e^{-\frac{\left(ln\left(A_b\right)-m_{log}\right)^2}{2\,s_{log}^2}}
    \label{Lognormal Distribution Equation}
\end{equation}
    
\begin{figure}[ht]
    \centering
    \includegraphics[width=0.6\linewidth]{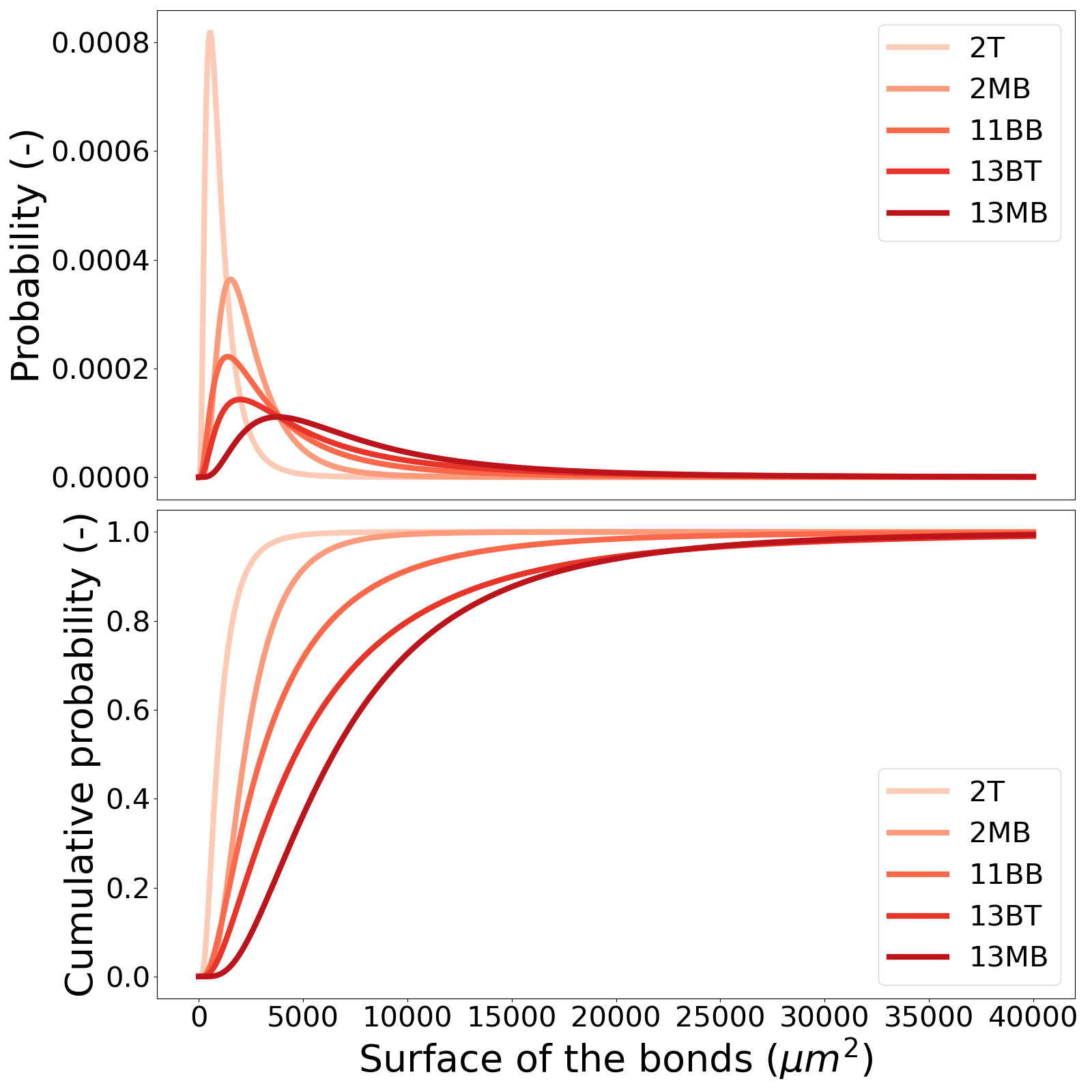}
    \caption{Examples of lognormal distributions used in this paper with the different cementations considered (see Table \ref{Parameters Used Cementation}).}
    \label{Lognormal Distribution Examples}
\end{figure}

It appears this distribution reproduces accurently experimental observations \cite{Sarkis2022}. Indeed, smaller bond surfaces and larger ones can be considered, both are crucial to the mechanical behavior of the sample. Examples of lognormal distribution used in this paper are given in Figure \ref{Lognormal Distribution Examples}.

%%=======================================================%%

\section{Bond breakage during the initialization}
\label{Bond Breakage Initialization Section}

Table \ref{Bond Breakage Initialization Table} shows the percentage of broken bonds (in comparison to the initial number of them) between the steps f (creation of the bonds) and h (loading at $P_{conf}$ and at $k_0^{init}$) of Figure \ref{Initial condition algorithm}. 
It is inevitable that some bonds will be broken as a consequence of the initialization algorithm, see Figure \ref{Initial condition algorithm}. 
Indeed, the objective is to generate a cemented granular material at a specified pressure $P_{cementation}$. Subsequently, the pressure increases to the confining value $P_{confinement}$. This increase is the source of some bond breakage (even prior to the test commencement). However, the discrepancy between the two pressures is crucial.
For the sake of argument, let us assume that $P_{cementation} = P_{confinement}$. The sample is generated, and the bonds are created. The sample is stable before the bond generation process, and thus, there are no unstable chain forces. The softening of the grains exerts a dominant influence on the evolution of the sample.
Moreover, the objective is to reproduce the experiments. In an oedometric test with acid injection (bond dissolution), the sample is initially cemented, set in the oedometer, loaded at confining pressure, and subsequently evolves with the injection of the acid (bond dissolution) \cite{Castellanza2004}. In the experimental context, the step of applying the confining pressure also results in bond breakage.

It appears Table \ref{Bond Breakage Initialization Table} there are more broken bonds for $k_0^{init}=0.2 \leq k_0^{attr}$ than for $k_0^{init}=0.7 \text{ or } 1.5 \geq k_0^{attr}$. During the initialization, the sample aims to reach the value $k_0 = k_0^{attr}$. In the case, $k_0^{init} = 0.2$ ($\leq k_0^{attr}$), the lateral dimension of the sample needs to extend. It is crucial to remember that confining pressure is still applied at the top wall. The combination of the two loadings induces mechanical bond breakage, see Equation \ref{Bond criteria}.

\begin{table}[h]
    \begin{tabular}{|l|c|c|c|}
        \hline
             Test characteristics & N initial & N final & Broken (\%) \\
        %\hline
        %     $P_{conf}=0.1$ MPa, $k_0^{init}=0.2$, 2T & 921 & 913 &  0.9\\
        %\hline
        %     $P_{conf}=0.1$ MPa, $k_0^{init}=0.7$, 2T & 963 & 962 & 0.1\\
        %     $P_{conf}=0.1$ MPa, $k_0^{init}=0.7$, 11BB & 7069 & 7068 &  0.01\\
        %\hline
        %     $P_{conf}=0.1$ MPa, $k_0^{init}=1.5$, 2T & 958 & 957 &  0.1\\
        %     $P_{conf}=0.1$ MPa, $k_0^{init}=1.5$, 11BB & 7086 & 7085 &  0.01\\
        %\hline
        \hline
             $P_{conf}=1$ MPa, $k_0^{init}=0.2$, 2T & 892 & 588 & 34\\
             $P_{conf}=1$ MPa, $k_0^{init}=0.2$, 2MB & 6192 & 5855 & 5\\
             $P_{conf}=1$ MPa, $k_0^{init}=0.2$, 11BB & 6681 & 6379 & 5\\
        %     $P_{conf}=1$ MPa, $k_0^{init}=0.2$, 13MB & 6810 & 6772 & 0.6\\
             $P_{conf}=1$ MPa, $k_0^{init}=0.2$, 13BT & 6614 & 6357 & 4\\
        \hline
             $P_{conf}=1$ MPa, $k_0^{init}=0.7$, 2T & 899 & 712 & 21\\
             $P_{conf}=1$ MPa, $k_0^{init}=0.7$, 2MB & 6185 & 5928 & 4\\
             $P_{conf}=1$ MPa, $k_0^{init}=0.7$, 11BB & 6802 & 6472 & 5\\
        %     $P_{conf}=1$ MPa, $k_0^{init}=0.7$, 13MB & 6682 & 6623 & 0.9\\
             $P_{conf}=1$ MPa, $k_0^{init}=0.7$, 13BT & 6744 & 6476 & 4\\
         \hline
             $P_{conf}=1$ MPa, $k_0^{init}=1.5$, 2T & 944 & 784 & 17\\
             $P_{conf}=1$ MPa, $k_0^{init}=1.5$, 2MB & 6104 & 5754 & 6\\
             $P_{conf}=1$ MPa, $k_0^{init}=1.5$, 11BB & 6709 & 6258 & 7\\
             $P_{conf}=1$ MPa, $k_0^{init}=1.5$, 13MB & 6664 & 6590 & 1\\
             $P_{conf}=1$ MPa, $k_0^{init}=1.5$, 13BT & 6817 & 6635 & 3\\
         \hline
         \hline
             $P_{conf}=10$ MPa, $k_0^{init}=0.2$, 11BB & 6973 & 3297 & 53\\
             $P_{conf}=10$ MPa, $k_0^{init}=0.2$, 13MB & 6867 & 3855 & 44\\
             $P_{conf}=10$ MPa, $k_0^{init}=0.2$, 13BT & 6988 & 4041 & 42\\
        \hline
             $P_{conf}=10$ MPa, $k_0^{init}=0.7$, 2T & 945 & 423 & 55\\
             $P_{conf}=10$ MPa, $k_0^{init}=0.7$, 2MB & 6487 & 3659 & 44\\
             $P_{conf}=10$ MPa, $k_0^{init}=0.7$, 11BB & 6847 & 3977 & 42\\
             $P_{conf}=10$ MPa, $k_0^{init}=0.7$, 13MB & 6986 & 5218 & 25\\
             $P_{conf}=10$ MPa, $k_0^{init}=0.7$, 13BT & 7079 & 4889 & 31\\
         \hline
             $P_{conf}=10$ MPa, $k_0^{init}=1.5$, 2T & 913 & 392 & 57\\
             $P_{conf}=10$ MPa, $k_0^{init}=1.5$, 2MB & 6438 & 3205 & 50\\
             $P_{conf}=10$ MPa, $k_0^{init}=1.5$, 11BB & 6886 & 3924 & 43\\
             $P_{conf}=10$ MPa, $k_0^{init}=1.5$, 13MB & 6888 & 4565 & 34\\
             $P_{conf}=10$ MPa, $k_0^{init}=1.5$, 13BT & 7084 & 4347 & 39\\
        \hline
    \end{tabular}
    \caption{Percentage of bonds broken during the initialization (steps f to h of Figure \ref{Initial condition algorithm}). Only the tests with a percentage $\geq 1\%$ are listed.}
    \label{Bond Breakage Initialization Table}
\end{table}

%%=======================================================%%

\end{document}